\documentclass[preprint2]{aastex}

\usepackage{xspace}
\usepackage{tipa}

\newcommand{\mic}{$\mu$m\xspace}
\newcommand{\Teff}{T$_\mathrm{eff}$\xspace}
\newcommand{\LaliLpah}{$L_\mathrm{ali}/L_\mathrm{PAH}$\xspace}
\newcommand{\up}{\textglobrise\xspace}
\newcommand{\down}{\textglobfall\xspace}

\shorttitle{Hydrocarbons in Herbig Ae systems}
\shortauthors{Acke et~al.}

\begin{document}


\title{Spitzer's view on aromatic and aliphatic \\ 
     hydrocarbon emission in Herbig Ae stars}


%
%
%
%
%
%
%

\author{B.~Acke\altaffilmark{1,2}, J.~Bouwman\altaffilmark{3},
  A.~Juh{\'a}sz\altaffilmark{3}, Th.~Henning\altaffilmark{3},
  M.E.~van~den~Ancker\altaffilmark{4}, G.~Meeus\altaffilmark{5},
  A.G.G.M.~Tielens\altaffilmark{6} and
  L.B.F.M.~Waters\altaffilmark{1,7}} 

\altaffiltext{1}{Instituut voor Sterrenkunde, K.U.Leuven,
  Celestijnenlaan 200D, B-3001 Leuven, Belgium}
\email{bram@ster.kuleuven.be}

\altaffiltext{2}{Postdoctoral Fellow of the Fund for Scientific
  Research, Flanders.}

\altaffiltext{3}{Max-Planck-Institute for Astronomy, K{\"o}nigstuhl
  17, D-69117 Heidelberg, Germany}

\altaffiltext{4}{European Southern Observatory,
  Karl-Schwarzschild-Strasse 2, D-85748 Garching bei M{\"u}nchen,
  Germany}

\altaffiltext{5}{Universidad Aut{\'o}noma de Madrid, Departamento de
  F{\'i}sica Te{\'o}rica C-XV, 28049 Madrid, Spain}

\altaffiltext{6}{Leiden Observatory, P.O. Box 9513, NL-2300 RA Leiden, the
  Netherlands}

\altaffiltext{7}{Astronomical Institute ``Anton Pannekoek'',
  University of Amsterdam, Science Park 904, 1098 XH Amsterdam, the
  Netherlands}


\begin{abstract}
The chemistry of astronomical hydrocarbons, responsible for the
well-known infrared emission features detected in a wide variety of
targets, remains enigmatic. Here we focus
on the group of young intermediate-mass Herbig Ae stars. 
We have analyzed the aliphatic and polycyclic aromatic hydrocarbon
(PAH) emission features in the infrared spectra of a sample of 53
Herbig Ae stars, obtained with the Infrared Spectrograph aboard the
Spitzer Space Telescope. 
We confirm that the PAH-to-stellar luminosity ratio is higher in 
targets with a flared dust disk. However, a few sources with a
flattened dust disk still show relatively strong PAH
emission. Since PAH molecules trace the gas disk, this
indicates that gas disks may still be flared, while the dust disk has
settled due to grain growth. There are indications that the strength
of the 11.3-\mic feature also depends on dust disk structure, with
flattened disks being less bright in this feature. 
We confirm that the CC bond features at 6.2 and
7.8\,\mic shift to redder wavelengths with decreasing stellar
effective temperature. Moreover, we show that this redshift is
accompanied by a relative increase of aliphatic CH emission and a
decrease of the aromatic 8.6-\mic CH feature strength.
Cool stars in our sample are surrounded by hydrocarbons with a high
aliphatic/aromatic CH ratio and a low aromatic CH/CC ratio, and vice
versa for the hot stars. We conclude that, while the overall
hydrocarbon emission strength depends on the dust disk's geometry, the  
relative differences seen in the IR emission features in disks around
Herbig Ae stars are mainly due to chemical differences of the
hydrocarbon molecules induced by the stellar UV field. Strong
UV flux reduces the aliphatic component and emphasizes the
spectral signature of the aromatic molecules in the IR spectra.
\end{abstract}





\section{Introduction}



Herbig Ae stars are pre- and zero-age main-sequence stars of a few solar
masses. They are surrounded by a circumstellar disk, a remnant of the
star-formation process. \citet{meeus01} have classified the spectral
energy distributions (SEDs) of these targets into two
groups, which reflect the geometry of the dust disk. Group~I sources have
a strongly flared outer disk, while the disks in group~II sources are
flatter. The outer disk surface of the latter lies in the shadow of
the inner disk 
\citep{dullemond01,dullemond04,acke09}. Both theory
\citep{dullemond04b,meijer08} and observations
\citep{ackesubmm} indicate that this difference in disk geometry is
due to dust grain growth and subsequent settling to the midplane.
Also the smallest dust grains in the disk surface layer appear to show
evidence for sedimentation,
although the correlation between grain size and flaring could also be
due to a radial gradient in grain size \citep[][Juh{\'a}sz
et~al. 2010]{bouwman08,furlan09}.

The material content of these circumstellar disks has been primarily
studied based on infrared (IR) spectra obtained with ground-based
instruments and spectrographs aboard
satellite missions such as the Infrared Space Observatory
\citep[ISO,][]{kessler} and, more recently, the Spitzer Space Telescope
\citep{werner04}. IR spectroscopy probes the thermal emission of the
warm surface layer of the disk. Micrometer-sized silicate dust grains,
both amorphous and crystalline, are commonly detected both because of
their abundance and the fact that their optical properties give rise
to strong bands in this wavelength regime
\citep[see the review by][]{henning09}. The spectral fingerprint of
the silicates has been interpreted in terms of grain growth and
crystallisation of the smallest grains in the disk
\citep[e.g.,][]{bouwman01,ackeiso,vanboekel05,sargent09}.
Besides silicate bands, the IR spectra of
many Herbig Ae stars display strong emission features from polycyclic
aromatic hydrocarbon (PAH) molecules. In this paper, we focus on these
molecules. They trace the carbonaceous component of the circumstellar
matter.

Next to its detection in Herbig Ae stars, PAH emission is
observed in a multitude of objects and environments, including the
interstellar medium (ISM), T~Tauri stars, reflection
nebulae (RNe), and, occasionally, evolved stars.
The emission features are the result of IR fluorescence of
PAH molecules pumped by ultraviolet radiation. Ionized PAHs can also
be excited by optical and near-IR photons
\citep[e.g.,][]{li02,mattioda05a,mattioda05b}. Because of 
the stochastic nature of
the heating process, PAH emission can be seen even far from the
illuminating UV source. For a comprehensive review on interstellar
PAHs, we refer to \citet{tielens08}.

Astronomical PAH features come in different flavors and show a broad
variety in shape, wavelength position and relative intensity. It is
believed that this diversity has its origin in a different
chemical constitution and ionization state of the carriers. The wealth
of laboratory measurements
and theoretical computations of hydrocarbon IR spectra has deepened
our understanding of the emission features. While the features at
3.3\,\mic, 8.6\,\mic and in the {10--14-\mic} range are attributed to CH
stretching and bending modes, the features at 6.2\,\mic and
\mbox{7--9\,\mic} are due to CC 
stretching modes \citep{allamandola}. Ionization enhances the
strength of the CC modes (\mbox{6--9\,\mic}) relative to the CH modes
(\mbox{10--14\,\mic}).
The molecular size mainly affects the CC features, while
the changes in the CH modes are expected to be more subtle. The
8.6-\mic feature, attributed to the CH in-plane bending mode, is
suggested to be produced by large (N$_C \sim 100$) compact PAHs
only \citep{bauschlicher08}. The \mbox{7--9-\mic} complex consists of
several sub-bands, with the most prominent at 7.6 and 7.8\,\mic
produced by small and large ionized PAHs respectively
\citep{bauschlicher08,bauschlicher09}.



Despite our theoretical and experimental knowledge of
PAHs, none of the astronomical spectra can be satisfactorily
fitted with the available set of computed/laboratory
spectra. Many analyses therefore focus on observational
correlations. Features produced by CC modes correlate better with each
other than with CH-mode features, and vice versa. \citet{hony01} have
investigated the CH out-of-plane modes (10--14\,\mic) in a sample
of various astronomical targets. The 11.3-\mic feature is ascribed to
lone (solo) CH groups found in PAH molecules with long straight edges,
while the features at 12.0, 12.7, 13.5 and 14.2\,\mic are linked to
duo, trio, quartet and quintet CH modes, indicative of corners in the
molecular structure. 

\citet{peeters02} categorized the astronomical PAH spectra in 3
classes, reflecting the peak position of the major features
in the 6--9-\mic range. Class~A contains the targets with the bluest
features, mainly general ISM sources, H{\sc ii} regions and RNe. Class~C,
characterized by the reddest 
features, consists mainly of cool carbon-rich (post-)AGB and red giant
stars \citep{sloan07,gielen09,smolders10}; Class~B sources have
intermediate 
feature peak positions, and include the Herbig Ae stars and most
PNe. \citet{sloan05} and \citet{sloan08} propose to categorize the
PAH spectra of Herbig Ae stars in a new class (dubbed
Class~B$^\prime$). Their 7--9-\mic features show an extended red wing
beyond 8.0\,\mic, which is absent in the Class~B features of PNe.

The wavelength shift of the features has been observationally linked
to the effective temperature of the central star
\citep{sloan07,boersma08,keller08}, with cooler stars
displaying redshifted IR 
emission features. This indicates that the chemistry of the emitting
hydrocarbon molecules depends on the properties 
of the stellar radiation field.

A plateau with superimposed narrow features has been detected in the
15--21-\mic range in a number of targets
\citep{beintema96,vankerckhoven00,sturm00,werner04b,sellgren07}. It is
our observation that most of these targets, if not all, belong to
Class~A. \citet{peeters04} attribute the narrow features to 
out-of-plane skeletal CCC modes of gas-phase PAH molecules. The
authors argue that the latter are large elongated PAHs with straight
edges and few pendant rings.

All features mentioned above are ascribed to PAHs. Their fundamental
constituents are aromatic (i.e.\ benzenoid) rings. Aliphatic
hydrocarbon molecules do not have such rings. The features 
at 6.8 and 7.2\,\mic, detected in a number of sources, are
attributed to aliphatic bonds (e.g.,
\citeauthor{furton99}~\citeyear{furton99};
\citeauthor{chiar00}~\citeyear{chiar00};
\citeauthor{dartois07}~\citeyear{dartois07}; however, see 
also \citeauthor{bauschlicher09}~\citeyear{bauschlicher09}).
The 6.8-\mic feature is linked to CH bending modes in both CH$_2$ and
CH$_3$ functional groups, the 7.2-\mic feature only to
CH$_3$. Hydrocarbon molecules with mixed hybridization display
aromatic as well as aliphatic emission features.

The work presented in this paper is performed in the framework of an
infrared 
study of a large sample of Herbig Ae stars, based on IR spectra
collected with the Spitzer Space Telescope.
\citet{acke09} investigated the geometry of the dust disk
based on the strength and slope of the thermal infrared emission. We
refine the interpretation of the Meeus group~I/group~II classification
and show that the height of the inner disk varies over the
sample. This leads to a varying amount of shadowing of the outer disk
and a change in the shape of the infrared continuum emission.
Juh{\'a}sz et~al. (2010, submitted to ApJ; hereafter J10) present the
full spectral decomposition into dust and continuum components. The
characteristics of the amorphous 
and crystalline silicate content of the upper disk layers are
related to stellar and disk properties.
The current paper focuses on the IR emission
features produced by aromatic and aliphatic hydrocarbon molecules.
Given their importance in the surface layers of
circumstellar disks, where they increase the gas temperature
through photoelectric heating and provide the surface for chemical
reactions, a better characterization of their chemical properties is
needed to improve our understanding of circumstellar disk structure and
evolution. Furthermore, the life cycle of hydrocarbon molecules in
space is poorly understood \citep[e.g.,][]{henning98b}. Here, we aim
to contribute to these topics. 


\section{Data set  \label{DR}}

\subsection{Spitzer IRS spectra}

We have gathered infrared spectra of 53 isolated Herbig Ae stars
obtained with the Infrared Spectrograph \citep[IRS,][]{houck04} aboard
the Spitzer Space Telescope. Most of these sources were observed
within the programs {\em The mineralogy of proto-planetary disks
  surrounding Herbig Ae/Be stars} (PID 3470, PI J.~Bouwman) and {\em
  Probing the disk mineralogy and geometry of Herbig Ae/Be stars} (PID
20308, PI B.~Acke). The spectra are 
characterized by a very high signal-to-noise ratio, typically several
hundreds. For a detailed description of the data reduction process,
the reduced spectra, and the sample targets, we refer to J10.

For each of the sample targets, published photometric data are gathered
to compile the spectral energy distributions. A reddened
\citet{kurucz} model was fitted to the observed UV-optical
photometry. The circumstellar IR excess flux is characterized by
several parameters such as excess magnitudes and spectral indices in
the near-to-far-IR and at (sub-)mm
wavelengths. Table~\ref{table_sed_parameters} summarizes a few of these
parameters.
Based on the SED, we have classified the sample into the Meeus groups
according to the method proposed by \citet{vanboekel05}. For details,
we refer to J10.

\subsection{Extraction of the PAH spectrum \label{extractPAH}}

In this section, we describe how the hydrocarbon emission features are
separated 
from the thermal emission of the circumstellar dust. The IRS
spectrum was split into two wavelength intervals (5--7 and 7--14\,\mic)
which are treated differently.

The 5--7-\mic wavelength range is characterized by the presence of
carbonaceous features on top of a relatively smooth dust continuum. In
this interval, the continuum was approximated with a single spline
through anchor points at 5.35, 5.45, 5.58, 5.82, 6.66, 7.06, 7.37,
7.40, 7.55 and 7.70\,\mic. Note that the broad \mbox{7--9-\mic} PAH
feature is treated as a continuum to the weaker features that are
located on top of
it. The features in the continuum-subtracted spectrum are
measured according to the procedure described in
Sect.~\ref{observables}.

The PAH features in the 7--14-\mic wavelength range are
less easily extracted from the original spectrum, as the underlying
``continuum'' may contain strong features of amorphous and crystalline
silicate dust. \citet{keller08} show that spline fits are only
applicable to spectra with a high feature-to-continuum
contrast. In spectra with strong silicate bands, artificial features
are created. \citet{keller08} therefore only consider the strongest
PAH sources in their analysis. In an attempt to reach weak PAH
features, we follow a different approach.
Six of our sample targets have a spectrum characterized by strong
PAH emission and by the absence/weakness of silicate features. These
sources are HD\,34282, RR~Tau, HD\,97048, HD\,135344B, HD\,141569 and
HD\,169142. Their PAH profiles are representative
for those observed in the entire sample, covering the observed range
in feature peak position, shape and width.
A single spline was fitted through continuum anchor points at 5.52, 5.83,
6.66, 7.06, 9.54, 10.32, 13.08 and 13.85\,\mic. 
The spline-subtracted spectra were used as PAH template spectra. Each
template spectrum is cut into five separate 
``PAH feature complexes'' at 5.5--7.0, 7.0--8.2, 8.2--9.2, 10.5--12.3 and
12.2--13.5\,\mic. 
The procedure to obtain full 5--16-\mic spectral decomposition of the
IRS spectra makes use of these templates. For each of the PAH
complexes, the fitting 
routine picks one of the six template spectra, appropriately
rescaled. The best fit for a certain target
includes a PAH spectrum which is a scaled combination of five template
spectra. The inclusion of PAH template spectra prevents
the fitting routine from compensating for the PAH 
features using the dust species in the model. Hence, the results of
the dust fit should be more reliable. Moreover, after subtraction of the
dust+continuum part of the best spectral fit to the original spectrum,
the residual spectrum is the real PAH spectrum of the
target. Fig.~\ref{residual_spectra_1.ps} shows the
residual spectra. A detailed description of the dust modeling can be
found in J10.

The procedure works well for most spectra. However, some residuals of
the crystalline silicate features remain. This is due to a slight
mismatch in peak wavelength between the observations and the computed
silicate opacities (see J10 for details). The residuals hamper an
accurate extraction of the PAH features with low peak-to-continuum
ratios, especially in the 
\mbox{10--14-\mic} region. Fig.~\ref{HD35187_residuals.ps} shows the
resulting PAH spectra of two silicate-dominated targets. In the case
of HD\,35187, the procedure leads to an excellent 
result; for HD\,244604, the extraction method creates strong artifacts
in the 7--12-\mic region. The features in this region cannot be
detected with 3-sigma significance. 

The aromatic-CCC bending-mode feature is present in the 16--19-\mic
range of a handful of sources. A spline through anchor
points at 14.0, 14.8, 15.0, 15.4, 15.6, 18.1, 18.6, 19.3 and
19.5\,\mic, approximating the underlying continuum, is subtracted.

\subsection{Derivation of the observables \label{observables}}

We aim at investigating whether the aliphatic and
aromatic features observed in the spectra display the signature of
chemical diversity and evolution of their carriers. To do so, the
detected features are captured in a number of representative
parameters. We stay
as close as possible to the real measurements, and avoid the
use of approximate line profiles such as Gaussian curves. In this
section, we describe how the observables are obtained. 

For each feature, a wavelength interval indicating its blue and red
limits is defined based on the best-quality PAH-dominated spectra where
the feature is detected (HD\,34282, HD\,97048, HD\,169142 and RR~Tau). This
wavelength interval was visually checked for each individual target in
order to avoid false detections. To minimize the effect
of residuals of the continuum subtraction, either the spline or dust
model, a straight line which connects the blue and red limit is
subtracted from the spectrum. The integrated line flux is the
total integral of the resulting continuum-subtracted
feature. Furthermore, the centroid position of the feature (i.e.\ the
flux-weighted barycenter), the full width at half maximum (FWHM), the
peak position and the peak flux are determined. Uncertainties on the first
two parameters were deduced directly from the error bars on the
spectrum. For the FWHM, peak position and peak flux, the following
approach was taken: 100 spectra of the feature were simulated around
the original spectrum, assuming a Gaussian error distribution with a
sigma equal to the error in each spectral pixel. In each of these
spectra, we determine the FWHM, peak position and peak flux of the
feature. The standard deviation of these measurements is taken as the
error on the feature parameter determined from the original
spectrum.

To quantify the influence of
our choice of the feature wavelength limits on the measurements, the
procedure described above was repeated over a slightly larger
wavelength interval: the blue limit was shifted one spectral pixel to the
blue, and the red limit one pixel to the red. The difference between
the two values for each parameter was used to
estimate the error induced by the choice of wavelength limits.
The final error on our measurements is the square root of the
quadratic sum of the latter, systematic, error and the
observational error described above.

Some of the features are a blend of individual features.
When the spectral resolution allows it, the components of these complexes
are separated. The largest feature is considered to be part of the
underlying continuum of the smaller feature(s). In general, a spline
is fitted to the wing of the large feature on one side of the smaller
feature, and to the continuum (which is zero in the continuum-subtracted
spectra) on the other side. The residual spectrum is that of the smaller
feature, the larger feature is set equal to the spline in the region
where the small feature is located. The parameters of both features
are determined as described above. Fig.~\ref{separate_6micron.ps}
shows the procedure for the 6-\mic region. This method was also used
to separate the 
7--9\footnote{Because almost all our sample
  targets belong to Class~B$^\prime$,
  with the dominant sub-band at 7.8\,\mic, we hereafter
  refer to this feature as the 7.8-\mic feature.} and 8.6-\mic
features and the 10.6, 11.3, 12.0 and 12.7-\mic features
(Fig.~\ref{separate_8micron.ps}). Again, an a priori fixed set of
anchor points was used for all stars in the sample.

%

From the line fluxes of the detected aromatic features, i.e.\ all features
except those at 6.8 and 7.2\,\mic, a total PAH flux is
computed. The same is done for the aliphatic features. The ratio of
the aliphatic and aromatic line fluxes is equal to their luminosity
ratio \LaliLpah. We also compute
the feature-to-stellar luminosity ratio L$_f$/L$_\star$ for each
feature $f$ separately. The stellar luminosity is that of the Kurucz
model fitted to the SED.


\section{Analysis}

\subsection{Detected features}

A detected feature is defined as a feature with a 
continuum-subtracted peak flux exceeding the 3-sigma level.
Strong aromatic features are detected at the well-known 
wavelengths 6.2, 7.8, 8.6, 11.3 and 12.7\,\mic. Secondary
features are observed at 5.7, 6.0, 10.6, 12.0 and 13.5\,\mic. We did
not detect the feature at 14.3\,\mic, linked to the quintet CH
out-of-plane bending mode, in any of the sample sources. 
The 16--19-\mic complex is measured in five targets (see
Fig.~\ref{detected16micron.ps}). 
Aliphatic features are detected in a number of sources, at
6.8 and 7.2\,\mic. In total, 13 spectrally separated features are
identified.

In 37 of the 53 sample sources (70\%), one or more significant PAH
features have been detected. The detection rate is 16/20 (80\%) in
group~I, and 21/33 (64\%) in group~II. Aliphatic features are found
in 22 of the 40 sources for which a short-wavelength spectrum is
available (55\%). The aliphatic detection rate is 11/15 in group~I
(73\%) and 11/25 in group~II (44\%).

Table~\ref{sample_avg} gives an overview of the sample-average
parameters of the detected features. We indicate the
possible carriers of each of the features as mentioned in the
literature. 
Tables~\ref{table_6_lineflux} and \ref{table_12_lineflux} list the
measured line fluxes and upper limits. Table~\ref{table_LpahLstar}
gives the fractional feature-to-stellar luminosities and
PAH-to-aliphatic luminosities.

\subsection{Correlations}

We searched for correlations between the feature parameters, the
stellar parameters and the disk parameters derived from the 
SED.
To compare two parameters, the Kendall tau-rank correlation
coefficient was computed, which is a non-parametric statistic
\citep{kendall38}.
A correlation is marked statistically significant if its
p-value is less than 1\%, under the assumption that no correlation is
present, before {\em and} after removal of the outliers. Hence,
apparent correlations based on only a few data points are removed, as
well as those induced by outliers.

The sample targets are classified in the two Meeus
groups. In all figures presented in this paper, group~I sources are
the black dots, group~II the red dots. 
For all feature parameters, we checked whether their values are
significantly different in one group with respect to the other. We
used the non-parametric Wilcoxon rank-sum test for 
assessing whether two independent samples of observations come from
the same distribution \citep{mann47}.

\section{Results \label{results}}

\subsection{The chemical diversity of PAHs in Herbig Ae
  stars \label{chemdiff}}

The Peeters spectral classification is interpreted in terms
of chemical diversity of the hydrocarbon carriers of the
features. 
The CC stretching feature at 6.2\,\mic in astronomical spectra is shifted
with respect to its theoretical position at 6.4\,\mic. This is also
the case in our sample.
The shift has been ascribed to the inclusion of nitrogen atoms in the
PAH skeletal structure. The feature is increasingly bluer when 
the N atom is substituted deeper in the PAH molecule
\citep{hudgins05}. Also the anion-to-cation 
ratio may play a role \citep{bauschlicher09}. However, the peak
positions of the 6.2 and 7.8-\mic features correlate
(Fig.~\ref{PP_6.2_7.8.ps}). While the general blueshift of the feature
from 6.4 to 6.2\,\mic may still be attributed to nitrogen pollution, the
latter cannot account for the progressive shift of the 6.2 and
7.8-\mic features. \citet{pino08} have performed laboratory
measurements and show that a shift of the 6.2-\mic feature to longer
wavelengths points to a higher aliphatic/aromatic content of the
hydrocarbon mixture.

The feature-to-stellar luminosity ratios of almost all PAH features (at
5.7, 6.0, 6.2, 7.8, 8.6, 10.6, 11.3, 12.0 and 12.7\,\mic) correlate
with each other. This indicates that these PAH features appear 
together in the spectrum of a Herbig Ae star, and roughly scale
together in
strength. The relative strengths of the aliphatic and aromatic
features, on the other hand, are not connected.
We also find dozens of significant correlations of line flux
ratios. The
strongest correlation in our sample is between the 6.2 and 7.8-\mic
features, both linked to CC modes. Tight connections were also 
found between the 6.2 and 8.6-\mic, the 7.8 and 8.6-\mic, and the
11.3 and 12.7-\mic features. Fig.~\ref{LF6260_7860.ps} shows two
examples of such correlations. The 6.2, 7.8 and 8.6-\mic features,
attributed to ionized PAHs, are closely related to each other, as well
as the features in the 
11--14-\mic range, due to out-of-plane CH bending modes of neutral and
ionized PAHs. 

\citet{hony01} and \citet{keller08} discuss the link between the
6.2/11.3 and 12.7/11.3 line flux ratios. The first is interpreted as
a measure for the degree of ionization, the second as a measure
for the irregularity of the PAH's edge. The authors conclude that both
ratios point to the degree of processing of the
molecules. An increase in PAH molecular size also enhances the
11.3-\mic feature with respect to the 6.2 and 12.7-\mic features. We
confirm this correlation between line flux ratios (p-value
1\%). \citet{keller08} also find a decrease 
of the 12.7/11.3 ratio with decreasing stellar temperature. Our data
do not show a clear correlation for the entire sample, but a
correlation may be present for the group~I sources only
(Fig.~\ref{lfteff_12.7_11.3.ps}). In any case, it appears that the
12.7/11.3 ratio is different in group~I and II. We
come back to this issue in the Sect.~\ref{featvsdisk}.

\subsection{The 16$-$19-\mic feature}

Five stars (HD\,34282, HD\,36917, HD\,97048, HD\,100453 and
HD\,135344B) display 
an emission feature at 16--19\,\mic. The complex is attributed to the
out-of-plane skeletal modes of large elongated PAH molecules
\citep{peeters04}. The detection is remarkable because our sample
targets are Class~B$^\prime$ sources, while
this feature complex is mostly seen in ISM-like Class~A objects. 

The strength and shape of the
16--19-\mic feature is not correlated to those of other features at
shorter wavelengths. This is consistent with the interpretation that
the short- and long-wavelength features are produced by different PAH
populations. Four of the five targets with detected 16--19-\mic
emission have a very red SED: their 30/13.5 continuum flux ratios are
among the highest in the sample (see
Table~\ref{table_sed_parameters}). Possibly, these sources are 
still surrounded by an envelope or extended outer disk in which the
chemistry of the PAH molecules is governed by the tenuous but harsh
interstellar UV field rather than that of the central star \citep[see
also][]{boersma08}.

Alternatively, the detection or otherwise of the feature may be
due to a contrast effect. The continuum emission at 20\,\mic is
dominated by warm dust grains close to the central star, while the
feature is produced in the outer disk via stochastic heating. The
extreme 30/13.5 continuum flux ratio could indicate that, while the
outer disk is still strongly flared, the inner disk has started to
settle. An additional indication for this hypothesis could be the
absence of the 10-\mic silicate emission feature in HD\,34282,
HD\,97048, HD\,100453 and HD\,135344B\footnote{
On the other hand, HD\,37411, RR~Tau, HD\,141569 and HD\,169142 are
also PAH sources without a 10-\mic silicate feature, but their spectra
do not show the 16--19-\mic feature.}: when silicate dust  
coagulates and grows beyond several $\mu$ms, the emission bands in the
mid-IR range disappear.
The settling of the inner disk implies a reduction of the warm
continuum emission at 20\,\mic and hence an increased
feature-to-continuum contrast. Later on, when 
the outer disk eventually starts to settle as well, the strength of
the 16--19-\mic complex decreases together with the degree of
flaring.

\subsection{Hydrocarbon molecules and silicate grains \label{sil_pah}}

Based on the IR spectra, the properties of the (sub-)\mic-sized
silicate grains in the upper layers of the disk can be derived. J10
show that only a few dust species contribute to the emission-band
spectrum: amorphous silicates with 
olivine and pyroxene stoichiometry, the Mg-rich crystalline silicates
forsterite and enstatite, and silica. Different grain sizes (0.1, 2
and 5\,\mic) are included in the spectral fit. The fitting routine was
performed on two wavelength ranges, 5--17\,\mic and 17--35\,\mic, to
sample the hot dust close to the star and the cooler dust further out
and deeper in the disk. This yielded the mass fraction of each of
the species, for each grain size and in both intervals. For details,
we refer to J10.

We have taken the dust mass fractions and searched for
correlations with the hydrocarbon feature parameters derived
here. None were found with a p-value below 5\%, with one marked
exception. The FWHM of the 6.2-\mic feature appears 
to be strongly correlated to the mass fraction of the 0.1-\mic silica
grains, derived from the spectral fit to the 17--35\,\mic spectrum
(Fig.~\ref{Silica_FWHM6.2.ps}). Targets with a broad 6.2-\mic feature
have less small silica grains in the cold/outer disk. It is conceivable
that this is a statistical false-positive correlation out of the
$\sim$3000 investigated comparisons between silicate and PAH parameters. 
However, the correlation has a very low p-value ($1 \times
10^{-5}$). The probability that such a strong random correlation
would occur here is only 3\%. Also the fact that both 
parameters were measured in disjunct wavelength ranges in the
spectrum, strengthens our confidence in the reality of the 
correlation.

The
relation is unexpected and we can only speculate about its
origin. PAHs are stochastically heated and their emission comes 
from further out in the disk surface, where grains in thermal
equilibrium are much colder. It is likely that the PAHs producing the
6.2-\mic feature and the cold silica grains are located in the
same physical region. 
\citep{hudgins05} mention that peripheral oxygen groups can influence
the 6.2-\mic feature position. Perhaps the
cumulative effect of a population of PAH molecules with
such peripheral groups broadens the 6.2-\mic feature. The
observed correlation would then indicate that the 0.1-\mic silica
grains and the PAH molecules are chemically competing for oxygen
atoms: the oxygen is either locked in peripheral groups of PAH
molecules, or included in 0.1-\mic silica grains.

\subsection{Influence of the disk geometry \label{featvsdisk}}

We checked for differences between Meeus group~I and II
sources. This classification is based on the thermal infrared emission
seen in the SED, and thus indicates the geometry of the
{\em dust} disk. Group~I sources have a flared dust disk, while the
dust disk in the group~II members has flattened, probably as a result of
dust grain growth and sedimentation to the midplane. 

In a number of papers \citep{meeus01,ackeiso,habart}, it has been 
claimed that the strength of the PAH features depends on the disk
geometry. A flared disk has a larger illuminated surface. Group~I
sources are therefore expected to produce stronger PAH emission. However,
\citet{keller08} do not find this correlation in their sample of
Herbig Ae/Be stars. Here, we elaborate on this.

Making use of radiative transfer models, \citet{meijer08} demonstrate 
that the degree of flaring of the dust disk largely depends on one
parameter: 
the mass in sub-\mic-sized dust grains. A higher dust mass results in
a higher opacity at optical wavelengths, a warmer outer disk, and more
flaring. \citet{acke09} show in their Fig.~3 that the flux excess
above the stellar photosphere at 60\,\mic is a tracer for the
dust mass, and hence for the 
degree of flaring. Fig.~\ref{exc30vsexc60.ps} compares
the IRAS 60-\mic excess to the excess at 30\,\mic. The latter is
determined from the Spitzer IRS spectra and the Kurucz model
for the star\footnote{The excess magnitude is defined as
M = $2.5 \log F_\mathrm{tot}/F_\star$.}. A very strong
correlation 
is noted, which indicates that also the excess at 30\,\mic traces the
disk geometry. Because IRAS~60-\mic photometry is not available
for all sample sources, and to rely only on the Spitzer IRS
spectra, we use the excess at 30\,\mic in the following.

We investigate whether the luminosity of 
individual PAH features, relative to the stellar luminosity,
correlates with the excess at 30\,\mic. The former
is the fraction of stellar flux captured and re-emitted by the PAH
molecules. We
find that the feature-to-stellar luminosity of most of the features (5.7,
6.0, 6.2, 7.8, 8.6, 10.6, and 12.7\,\mic) increases significantly with
increasing 30-\mic excess. Fig.~\ref{PAH62vsEXC30.ps} shows the
correlation for the most frequently detected feature, at
6.2\,\mic. Also the 
total PAH-to-stellar luminosity ratio 
correlates with the far-IR excess (p-value $9 \times 10^{-5}$).

It is clear that the disk geometry influences the PAH-to-stellar
luminosity ratio. However, the large amount of scatter indicates that
the degree 
of flaring is not the only parameter. Source-to-source differences in
PAH abundance are certainly important.
Furthermore, there are targets which have a flattened
dust disk (group~II), but are still strong PAH emitters. A 
solution to this apparent discrepancy has been proposed in a few recent
papers. \citet{vanderplas08} and \citet{fedele08} show that the dust
disk can decouple from the gas disk, probably due to dust grain growth
and subsequent settling to the midplane. This results in a flattened 
dust disk, while the gas disk can still have a flared
structure. \citet{verhoeff10} have further elaborated
on this idea and performed detailed modeling of HD\,95881, a target with
an extremely blue group~II-type SED, but strong PAH emission. 

Along the same lines, \citet{dullemond07} have modeled dust
sedimentation in protoplanetary 
disks with PAHs. They find that, if the 
PAH abundance stays the same, sources with sedimented dust disks
should display {\em more} pronounced PAH features than
disks with perfect vertical mixing. Two effects are important: the
reduction of the continuum emission (increased contrast) and the
increased amount of UV-exposed PAH molecules (increased PAH
luminosity). Under the assumptions of the models, a decrease in far-IR
continuum emission 
should be accompanied by an increase of PAH luminosity. The opposite
trend is observed, however. \citet{dullemond07} speculate that a
significant fraction of the PAHs is removed through aggregation. In
any case, a reduction of PAH abundance in the flared gas disk above
the dust disk, e.g.\ through PAH photodestruction, is necessary to
reconcile theory and observations.

We found indications that there is indeed
a small difference in PAH chemistry between flared and flattened 
disks. There is a strong correlation between the peak flux ratios
11.3/6.2 and 
11.3/7.8 (Fig.~\ref{PF11_62and11_78.ps}). This is not surprising, as
the 6.2 and 7.8-\mic features are both CC modes and thus strongly
linked to each other (see 
Sect.~\ref{chemdiff}). From the figure, however, it seems that the
distributions of the 11.3/6.2 and the 11.3/7.8 ratios are shifted
towards lower values in 
group~II. Also the 12.7/11.3 line flux ratio appears to be higher in
this group ($0.29 \pm 0.08$ vs. $0.22 \pm 0.05$ in group~I, p-value
0.5\%). This suggests that the 11.3-\mic feature is weak in the
targets with a small far-IR excess, and may indicate that the PAH
molecules in targets with a flattened 
dust disk are more processed, i.e.\ rough-edged
and/or smaller, than those in flared disks.

\subsection{Aliphatic versus aromatic hydrocarbons \label{aliaro}}

Almost all stars in our sample
display Class~B$^\prime$ PAH features, but it
is clear that even within 
this class, spectral differences occur. Here we argue that these changes
are mainly due to the stellar UV field to which the hydrocarbon
molecules are exposed. 

The centroid (and peak) position of the 7.8-\mic feature
decreases with increasing effective
temperature. Fig.~\ref{centeff_7.8.ps} shows the data for our sample
of Herbig Ae stars. We have expanded the sample with a set of
pre-main-sequence 
stars of different stellar masses, obtained in other Spitzer IRS
programs. On the low-mass end, we have included five T~Tauri stars
with detected PAH emission. RX~J1842.9$-$3532,
RX~J1852.3$-$3700, HD\,143006, RX~J1612.6$-$1859A and
1RXS~J132207.2$-$693812 were observed within the {\em Formation and
Evolution of Planetary Systems} (FEPS) Legacy Science Program (PID
148, PI M.~Meyer). The
spectra and stellar effective temperatures were taken from
\citet{bouwman08}. On the high-mass end, six Herbig Be stars were
added. The spectra of HD\,290770, Hen~3$-$180, LkH$\alpha$~257, PDS~216,
PDS~344 and RNO~6 were acquired within the Spitzer Program {\em Structure
and composition of disks surrounding Herbig Be stars} (PID 50180, PI
M.~van~den~Ancker). Their effective temperature has been estimated
based on their spectral type. The correlation between 7.8-\mic
centroid position and \Teff is clear for the Herbig Ae stars, and is
even more prominent when including the lower- and higher-mass young
stars. The contribution of the red sub-bands appears to be more
significant in cooler stars than in hotter stars, which is supported
by the fact that the width of the 7.8-\mic feature increases with
redshift (Fig.~\ref{cen_vs_fwhm_78.ps}). Finally, also the peak and
centroid position of the 6.2-\mic feature are correlated to the
stellar effective temperature (p-value $5 \times 10^{-4}$ and $4
\times 10^{-3}$ resp.): cool stars display red features.

The correlation between \Teff and 7.8-\mic wavelength position was
already found in smaller samples by 
\citet{sloan07}, \citet{boersma08} and \citet{keller08}. As we have
shown, it holds for a broad variety of 
pre-main-sequence stars, with spectral types ranging from M0 to early
B. \citet{smolders10} have recently extended the  
correlation to lower \Teff and redder features with the detection of
hydrocarbon emission in S-type AGB stars.
The stellar radiation field strongly influences the chemistry of the
surrounding hydrocarbon molecules, 
regardless of the object's evolutionary status or specificities of
the circumstellar environment.

\citet{sloan07} propose that a
red feature indicates the presence of a hydrocarbon mixture with a
higher aliphatic/aromatic ratio, an idea that is supported by laboratory
measurements \citep{pino08}. Here we report a few newly found correlations
in our sample of Herbig Ae stars that provide further evidence for this
interpretation.


The position and width of the 7.8-\mic feature are linked
to the effective temperature of the central star. Moreover, the
strength of the aromatic 8.6-\mic CH in-plane bending mode, relative
to the 6.2 and 7.8-\mic CC features, and the aliphatic-to-aromatic
luminosity ratio\footnote{This is the ratio of the summed line fluxes
  of the aliphatic 6.8 and 7.2-\mic features with respect to those of
  the detected PAH features in the covered wavelength range.} depend on
\Teff as well. In fact, all these parameters 
are interconnected. Table~\ref{table_corr} summarizes these
correlations. Figs.~\ref{cen78_vs_LF86_78.ps} and
\ref{LaliLpahvscen78.ps} show that, while targets with a red
7.8-\mic feature (i.e.\ cool stars) have a low flux ratio of the
aromatic 8.6-\mic CH and 7.8-\mic CC features, they produce
fractionally more aliphatic emission.
It is clear that illuminated 
hydrocarbon molecules around stars with different effective
temperature are chemically different. 

Herbig Ae stars are by nature strong photospheric UV emitters. Because
of the rather low accretion rates in most targets, the additional
UV excess due to accretion energy dissipation is relatively
unimportant for these stars. We can therefore consider the effective
temperature as a direct indicator of the stellar UV radiation field.

Strong UV radiation appears to preferentially destroy or convert the
aliphatic component in the disks. Photodestruction of the CH bonds is
expected from laboratory experiments
\citep[e.g.,][]{munoz_caro01}. This effect 
is stronger for aliphatic molecules, as aromatic rings give more
stability to the molecule by allowing distribution of the UV-induced
excitation energy. This is consistent with the lower line fluxes of 
the 8.6-\mic feature relative to the aliphatic CH features observed
in cooler stars. The disappearance of the 6.8 and 7.2-\mic features,
both 
linked to CH bonds in aliphatic hydrocarbons, could therefore point to
a more rapid dehydrogenation of the aliphatic molecules, rather than
a complete removal of this component. However, the experiments of
\citet{munoz_caro01} show that the aromatic CH bonds are
also eventually destroyed by the UV radiation. The
aromatic feature at 8.6\,\mic should therefore become weaker
with increasing UV strength as well. The opposite trend is observed:
sources with strong UV fields display a higher aromatic CH/CC flux
ratio. We note that this interpretation relies on the identification
of the 8.6-\mic feature as due to aromatic CH modes. Alternatively,
it may be that also aromatic CC modes contribute to its strength. In
the latter case, the observations show that the aromatic hydrocarbons
undergo structural changes induced by UV field, resulting in Class~A
features and an increased contribution of CC modes to the 8.6-\mic
feature.

The observed increase in the aromatic CH strength at higher
temperatures is consistent with experimental work by
\citet{mennella01}, who suggest that the heating of carbon grains
should reduce the number of aliphatic CH bonds and increase the
number of aromatic CH bonds. Also UV irradiance increase the relative
degree of aromatic clustering, but in thermal annealing at
temperatures above 300\,K, this increase is accompanied with a growth
of the graphitic clusters. Under UV irradiation, the clusters remain
small \citep{mennella98}. 

In diffuse interstellar clouds, large ($>30$~C atoms) PAHs
are expected to have normal hydrogen coverage (i.e.\ one H atom per
peripheral C atom; aromatic bond), while very large PAHs may be fully
hydrogenated \citep[i.e.\ two H atoms; aliphatic bond,][]{lepage03}. 
The transition of full to normal hydrogen coverage could occur through
UV photolysis of the aliphatic CH bonds and replacement with aromatic CH
bonds. This would reduce the aliphatic fingerprint in the
infrared spectrum and pump the aromatic CH/CC ratio. Because UV
processing keeps the cluster sizes small, thermal annealing may be
needed to grow the clusters and increase the aromatic CH emission
strength. The
observational trends in our sample then indicate that not
only UV photodestruction, but also thermal annealing of carbon grains
and hydrogenation in the UV-shielded disk interior, is important in
circumstellar disks around Herbig Ae stars.


\section{Conclusions \label{conclusions}}

In this paper, we have investigated hydrocarbon emission in the
Spitzer IRS spectra of a sample of 53 Herbig Ae stars. We find that
70\% of the sources display PAH emission, and 55\% aliphatic
features at 6.8 and/or 7.2\,\mic. Five targets (9\%)
display a feature complex at 16--19\,\mic. Below we list our main
conclusions.

\begin{itemize}
\item We confirm the correlation between the 
PAH feature-to-stellar luminosity ratio and the shape of the infrared
spectral energy distribution, first mentioned by
\citet{meeus01}. PAHs are a tracer of the gas disk, while the SED
probes the thermal dust emission. If the dust and gas disk are
colocated, the trend indicates that the illuminated surface of a flared 
disk is larger than that of a flattened disk. The volume occupied by
excited PAHs is therefore larger in the first group.
It is suggested in the literature that sources with settled dust
disks can nonetheless produce strong PAH emission if their gas disk is
still flared and exposed to stellar UV flux. This could explain the
existence of the few targets with flattened dust disks but strong PAH
emission. However, most targets follow the general trend. We conclude
that PAHs in the dust-poor disk atmosphere are either rapidly removed
from the gas phase, or that the gas and dust in most disks are well
coupled and settle together.


\item The stellar radiation field determines the chemistry of the
hydrocarbon population in the circumstellar disk. Ranked according to
increasing stellar effective temperature, the infrared spectra of
Herbig Ae stars show an increase of the aromatic CH emission, but a
decrease of the aliphatic CH emission. Aliphatic CH bonds are more
easily destroyed than aromatic CH bonds, which explains the decrease
in aliphatic signature. However, this effect cannot
explain the increase in the aromatic CH/CC ratio with increasing UV
strength. The solution may be found in thermal annealing and the
hydrogenation balance of the PAHs. The latter can only influence
the PAH chemistry if turbulent vertical mixing is important at
the disk's surface, and brings PAH molecules from UV-immersed regions
to the disk interior and back.

\item
  The 6.2 and 7.8-\mic CC features are redder in cooler stars. The link
  with the relative strength of the aliphatic features indicates that
  the wavelength shift indeed is a measure for the aliphatic/aromatic 
  content ratio of the hydrocarbon mixture.




\item The strong connection between spectral fingerprint and stellar
  effective temperature shows that hydrocarbons immediately react
  to the UV field to which they are exposed. Therefore, the chemistry of
  hydrocarbon molecules 
  does not in general trace disk evolution. However, the observations
  indicate that targets with a flattened 
  dust disk harbor hydrocarbons which may be more processed. They
  appear to have rougher molecular edges and/or are smaller than those
  in flared dust disks.

\end{itemize}

We have expanded the knowledge of the chemistry of hydrocarbon
molecules in disks. Once irradiated, the free-flying molecules are 
quickly chemically altered by the stellar UV field and their history
is deleted. However, PAHs in disks must have a significant aliphatic
component when they are appear at the disk surface. Laboratory 
experiments have shown 
that the formation of aliphatic (CH$_2$) groups on the edges of PAH
molecules can already occur under very mild conditions \citep[T
$\approx$ 400\,K]{jaeger06}. Such conditions are available in the warm
regions of disks around Herbig Ae stars, which makes it probable that
in-situ production and (re-)hydrogenation of the
hydrocarbon mixture are processes which take place below the disk
surface.



\acknowledgments

This work is based on observations made with the Spitzer Space
Telescope, which is operated by the Jet Propulsion Laboratory,
California Institute of Technology under a contract with NASA.
BA thanks Kees Dullemond for helpful discussions.






\clearpage






\begin{figure*}
\centering
\includegraphics[height=0.9\textheight]{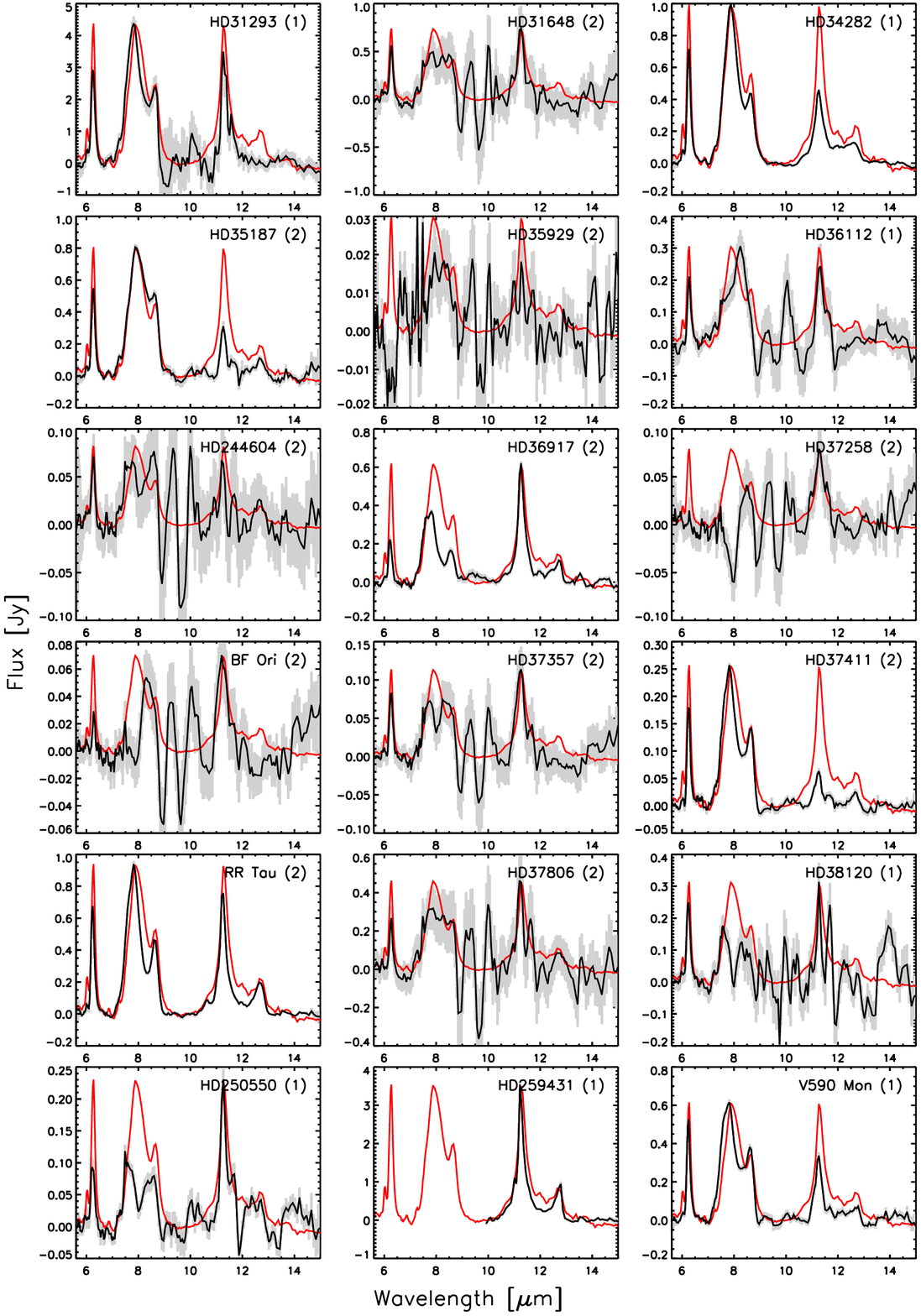}
\caption{Residual PAH spectra. Next to the source name, the Meeus
  group classification is indicated. As a reference, the PAH spectrum of
  HD\,169142 is overplotted in red, scaled to the peak of the residual
  spectrum.}
           \label{residual_spectra_1.ps}%
 \end{figure*}

\begin{figure*}
\centering
\includegraphics[height=0.9\textheight]{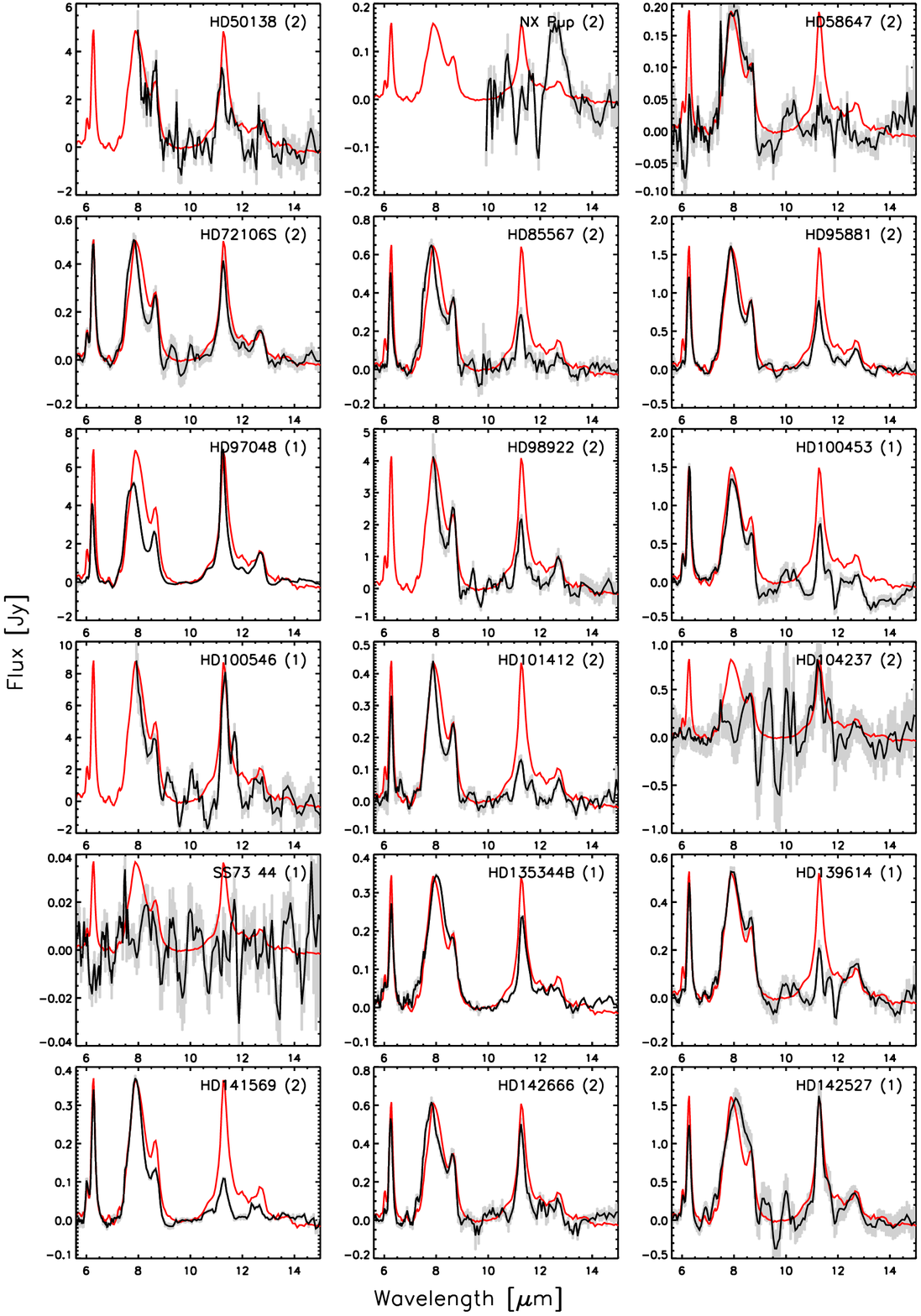}
\addtocounter{figure}{-1}
\caption{Cont'd.}
           \label{residual_spectra_2.ps}%
 \end{figure*}

\begin{figure*}
\addtocounter{figure}{-1}
\centering
\includegraphics[height=0.9\textheight]{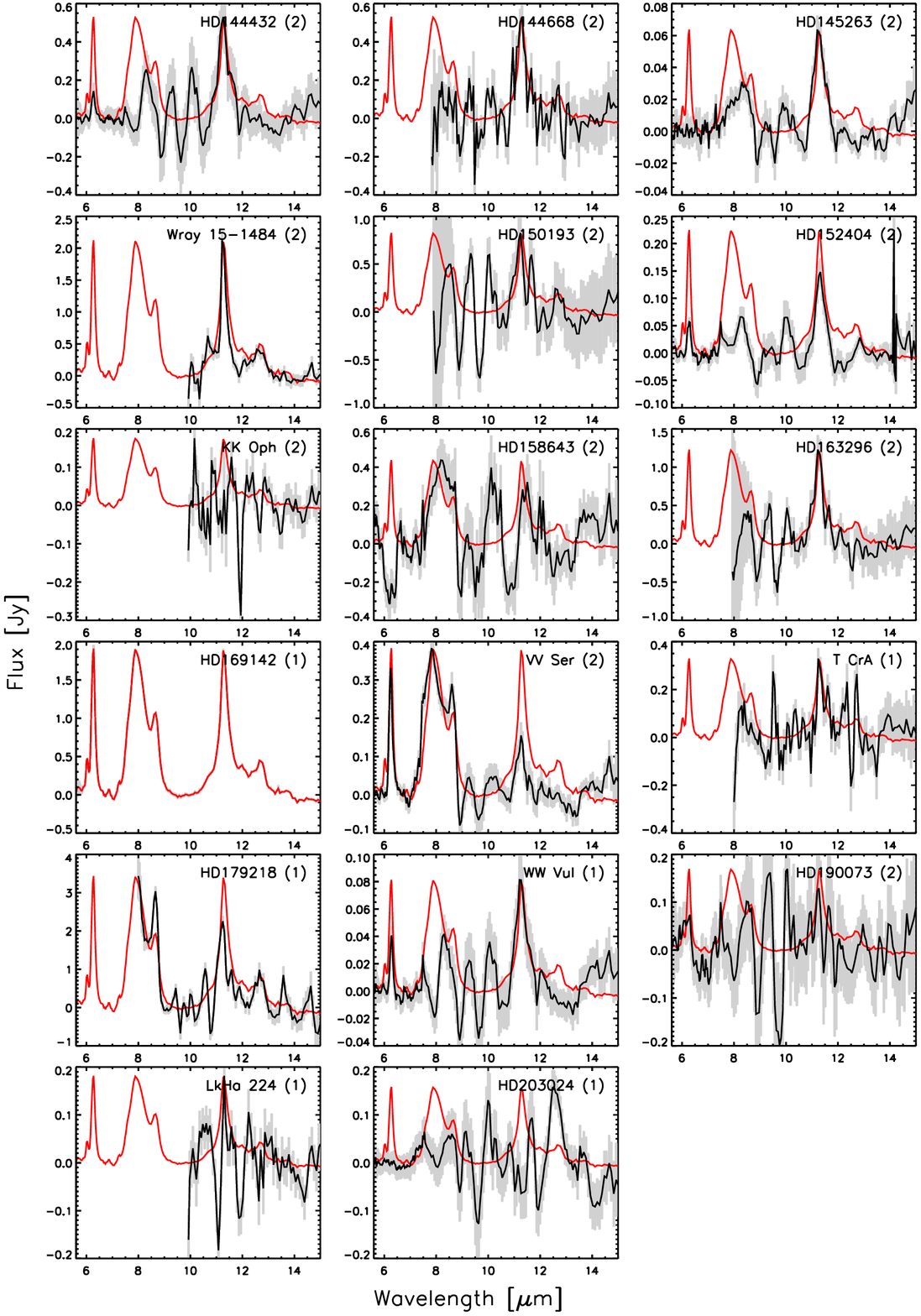}
\caption{Cont'd.}
           \label{residual_spectra_3.ps}%
 \end{figure*}

\begin{figure*}
\centering
\includegraphics[width=0.6\textwidth]{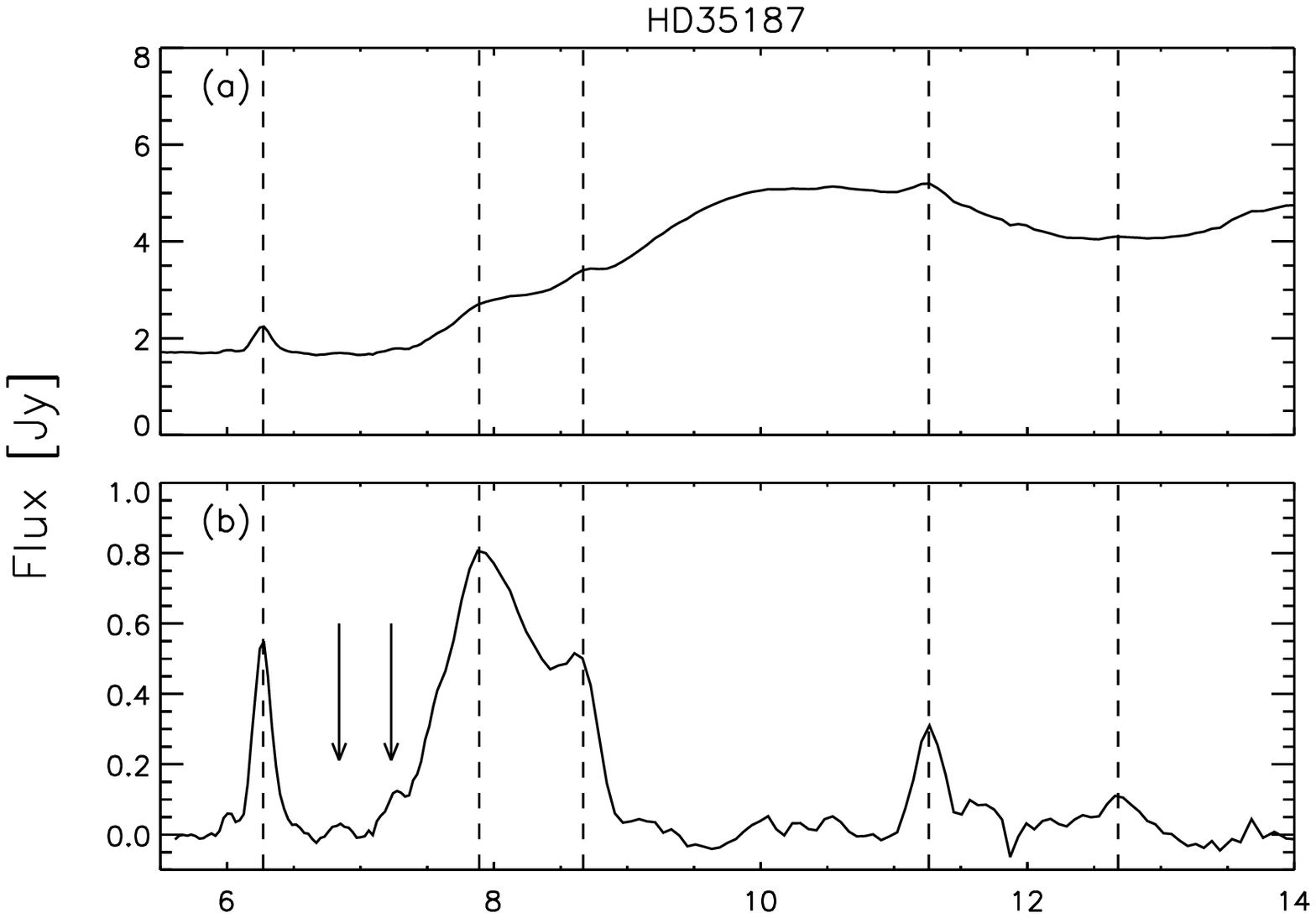} 
\includegraphics[width=0.6\textwidth]{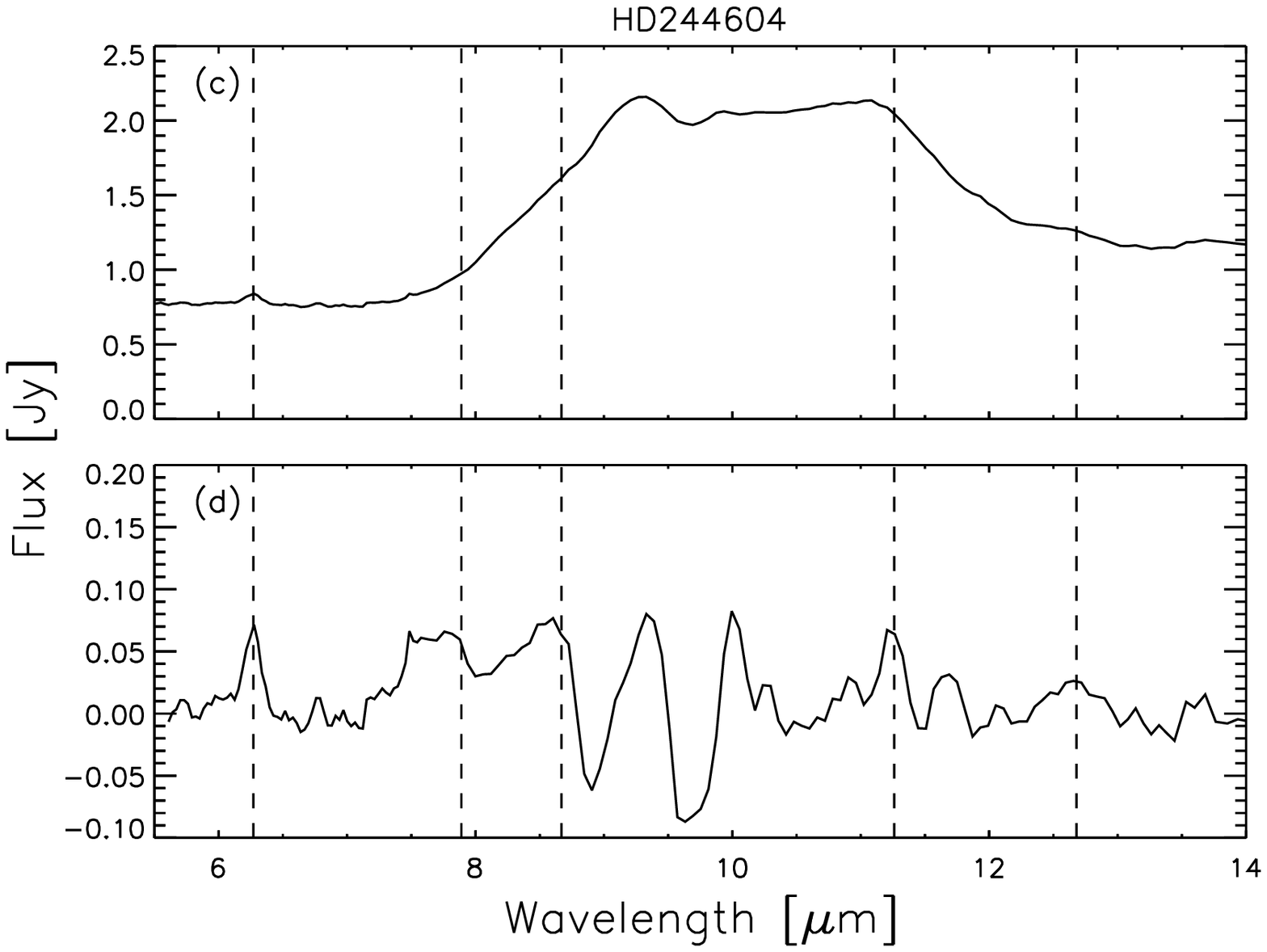}
\caption{ {\it (a)} Spitzer IRS spectrum of HD\,35187 and {\it (b)} PAH
  spectrum, extracted by subtracting the best fit to the
  dust and continuum (J10). The major features at 6.2, 7.8, 8.6,
  11.3 and 12.7\,\mic are clearly visible (dashed lines), as well as
  the aliphatic features at 6.8 and 7.2\,\mic (arrows). {\it (c)}
  Spitzer IRS spectrum of HD\,244604 and {\it (d)} PAH 
  spectrum. Residuals of the silicate dust fit
  are present on a few-percent level and hamper a reliable detection
  of the (weak) PAH features longward of 7\,\mic.}
           \label{HD35187_residuals.ps}%
 \end{figure*}

\begin{figure*}
\centering
\includegraphics[width=\textwidth]{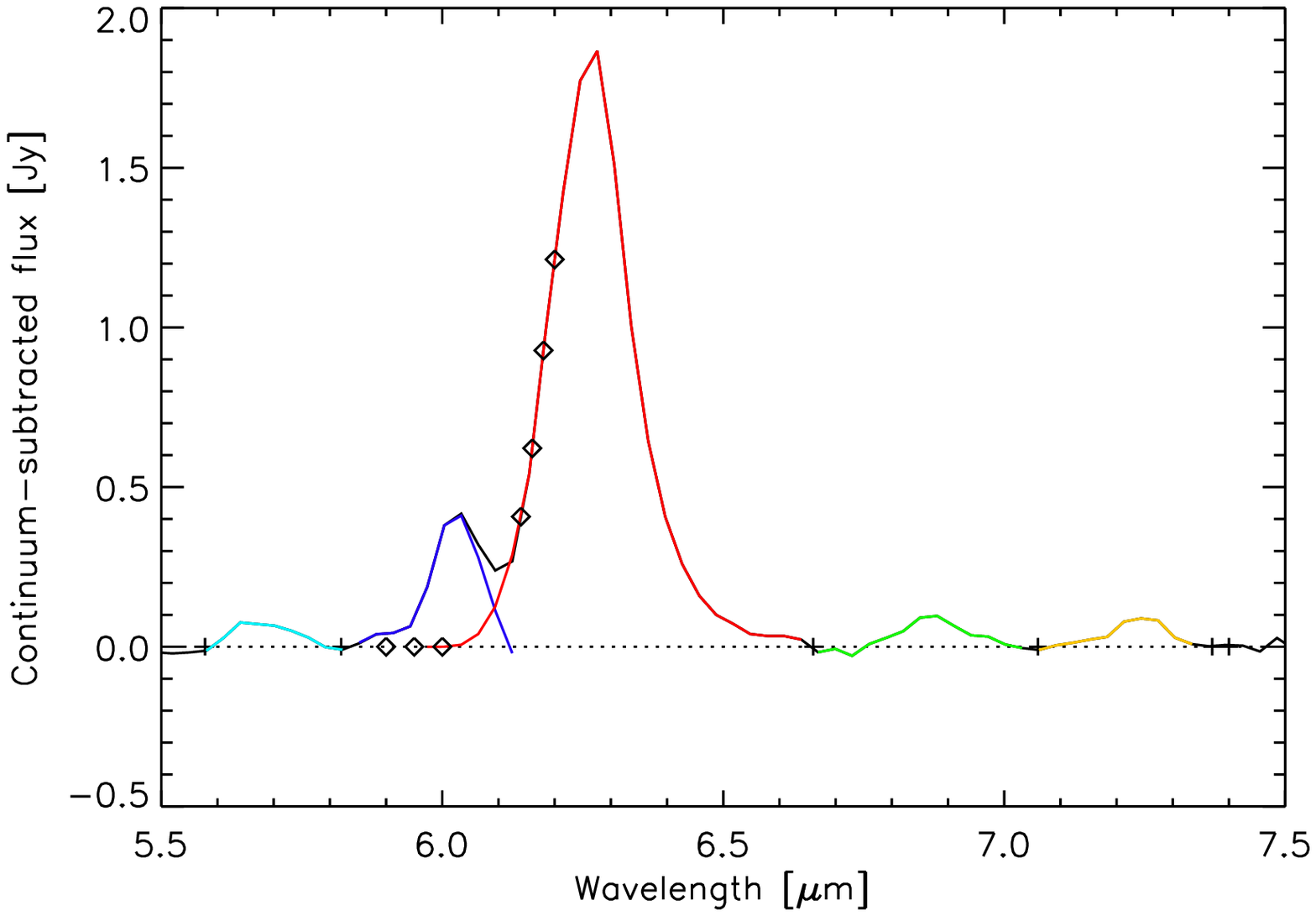}
\caption{Extraction of the features in the 6-\mic range of the 
  IRS spectrum of HD\,169142. A continuum spline, through the anchor
  points indicated with $+$, was subtracted. The 5.7, 6.8 and
  7.2-\mic features are not blended and can be measured directly.
  A spline is fitted 
  to the blue wing of the large 6.2-\mic feature at well-chosen
  wavelength positions (diamonds). It is assumed that the large
  feature reaches the continuum level (i.e.\ zero) at 6\,\mic. The
  resulting line profiles (colored lines) are plotted on top of the
  original spectrum (black). See text for details.}
           \label{separate_6micron.ps}%
 \end{figure*}

\clearpage

\begin{figure*}
\centering
\includegraphics[width=0.6\textwidth]{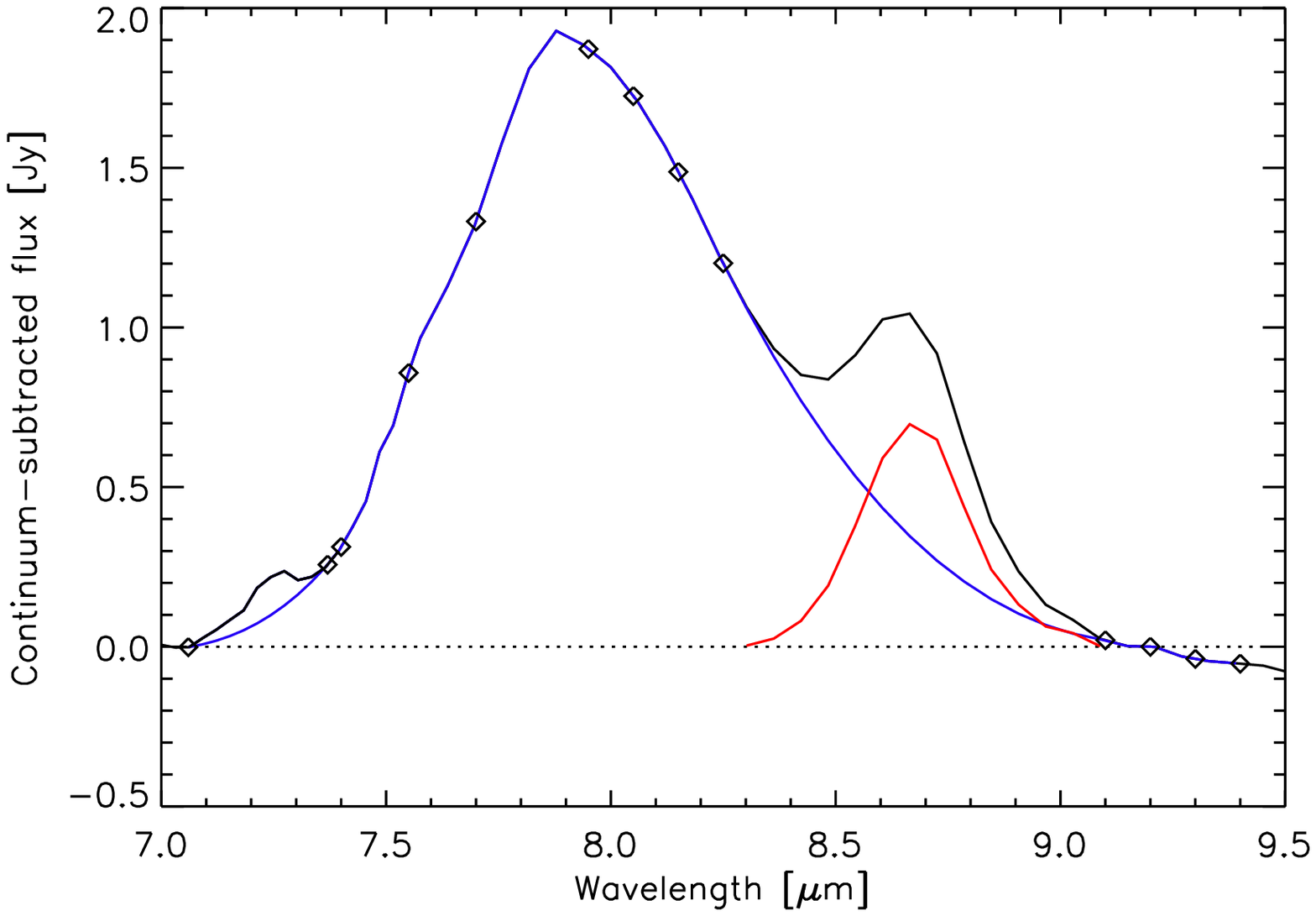}
\includegraphics[width=0.6\textwidth]{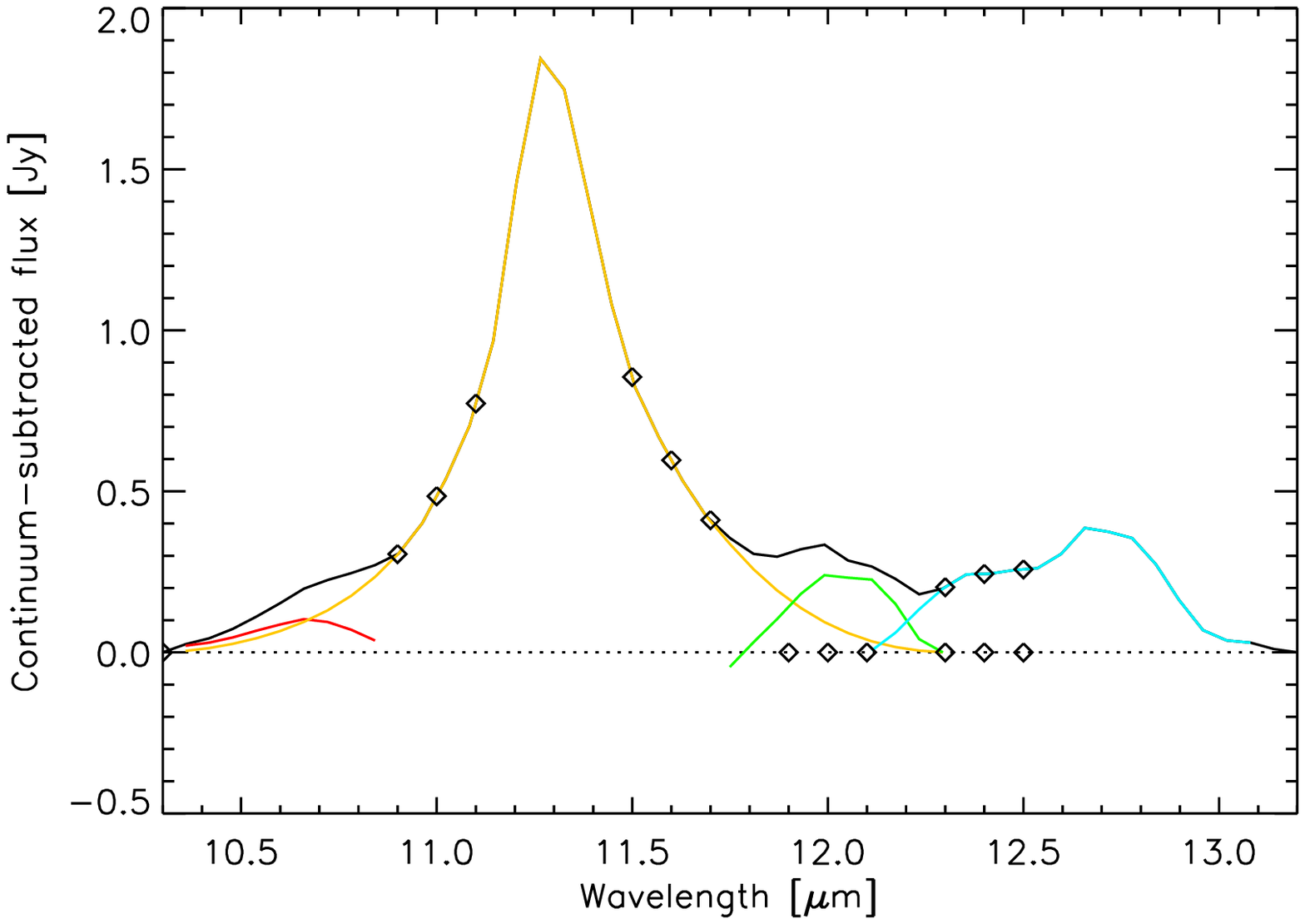}
\caption{Extraction of the features in the 7--9-\mic range ({\em
    top}) of the 
  residual PAH spectrum of HD\,169142. Splines
  are fitted at a priori fixed anchor wavelengths (diamonds). 
  The resulting line profiles (colored lines) are plotted on top of
  the original spectrum (black). Note that the 7.2-\mic feature is
  extracted, considering the 7.8-\mic feature as continuum emission
  (see also Fig.~\ref{separate_6micron.ps}). {\em Bottom:}
  Extraction of the features in the \mbox{10--13-\mic} region. }
           \label{separate_8micron.ps}%
 \end{figure*}

\begin{figure*}
\centering
\includegraphics[width=\textwidth]{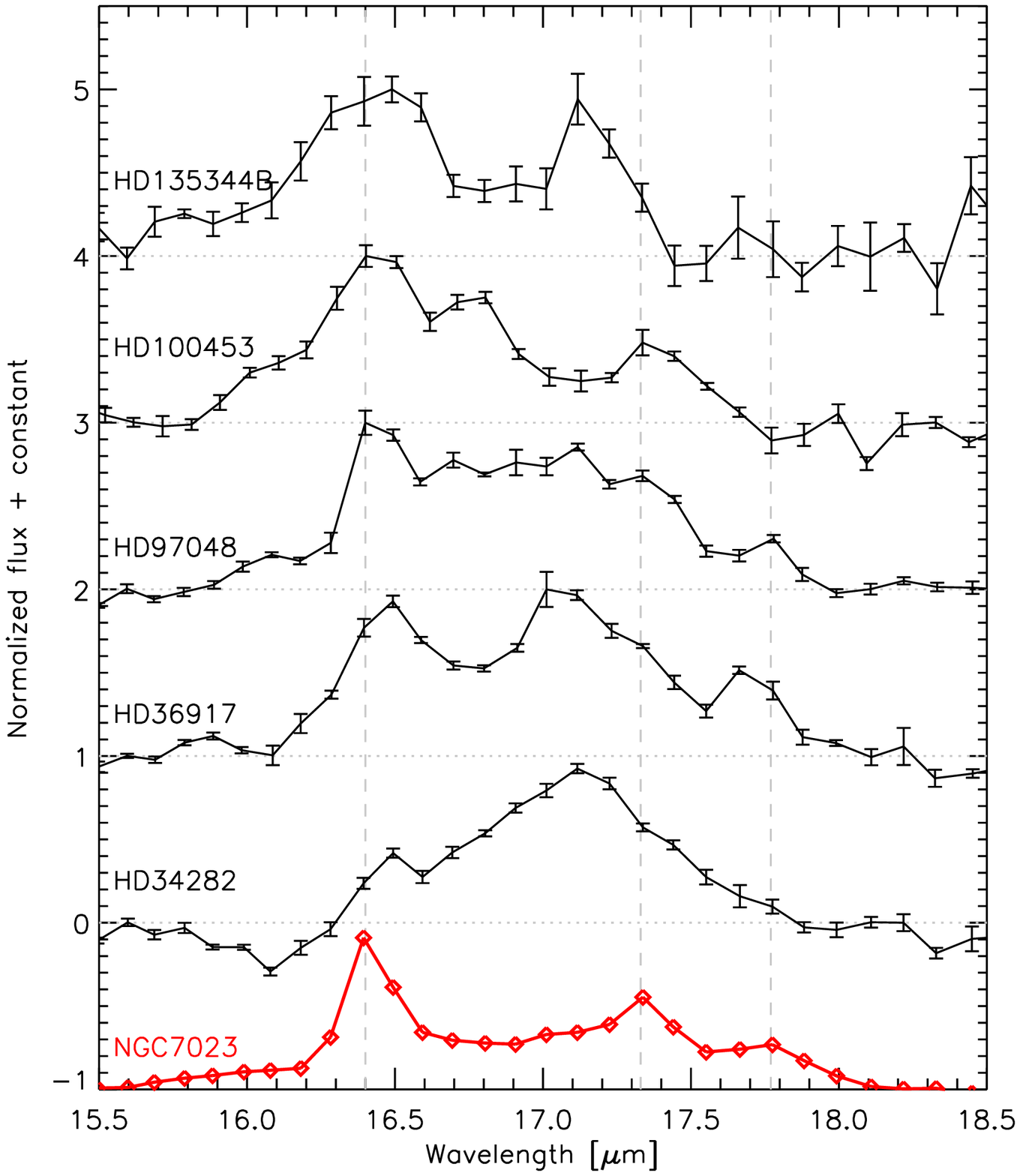}
\caption{Detected features of the skeletal CCC out-of-plane bending
  mode. The spectrum of the NGC~7023 reflection nebula
  is shown at the bottom as a reference \citep{sellgren07}. It is
  rebinned to a lower resolution of 160 to allow direct comparison
  with the Herbig Ae spectra. The full vertical line indicates the
  position of the H$_2$ 0-0 S(1) molecular line. Even if it would be
  present, confusion with the broad PAH feature can be excluded.}
           \label{detected16micron.ps}%
 \end{figure*}

\begin{figure*}
\centering
\includegraphics[width=\textwidth]{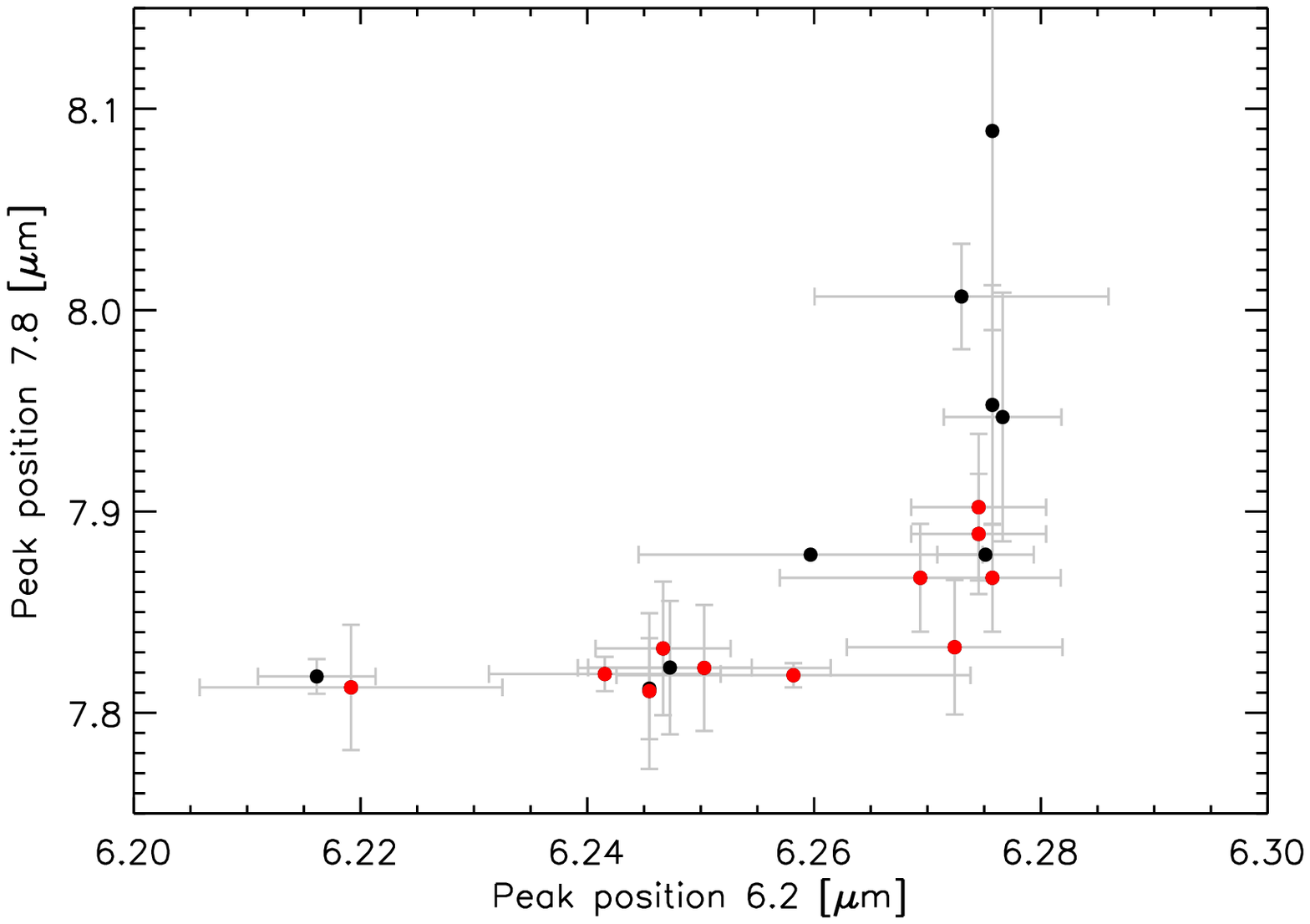}
\caption{The peak position of the 6.2 and 7.8-\mic features (p-value
  $3 \times 10^{-5}$). The black dots represent group~I sources, the red
  group~II. This convention is used throughout the entire paper.}
           \label{PP_6.2_7.8.ps}%
 \end{figure*}

\begin{figure*}
\centering
\includegraphics[width=0.6\textwidth]{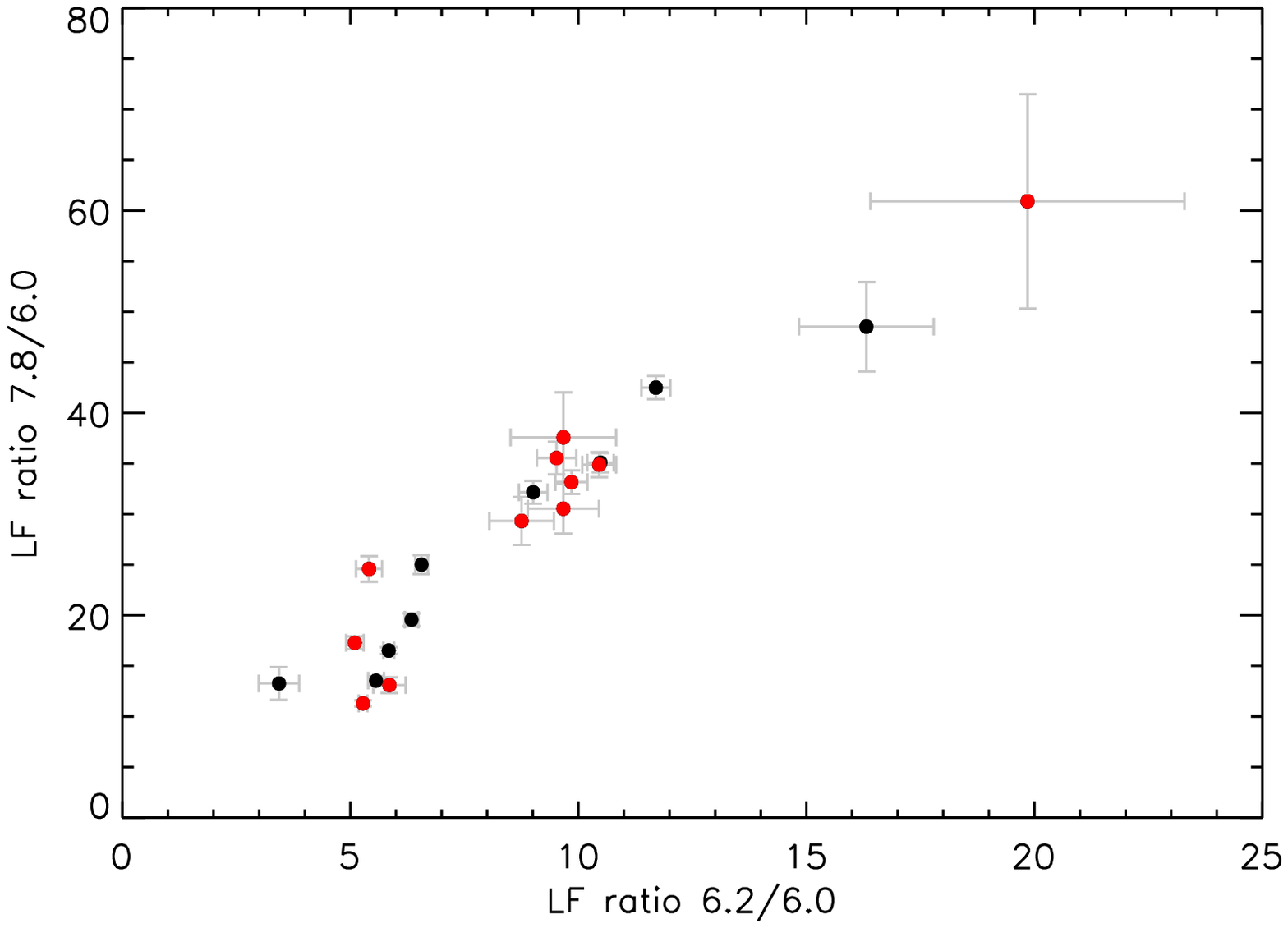}
\includegraphics[width=0.6\textwidth]{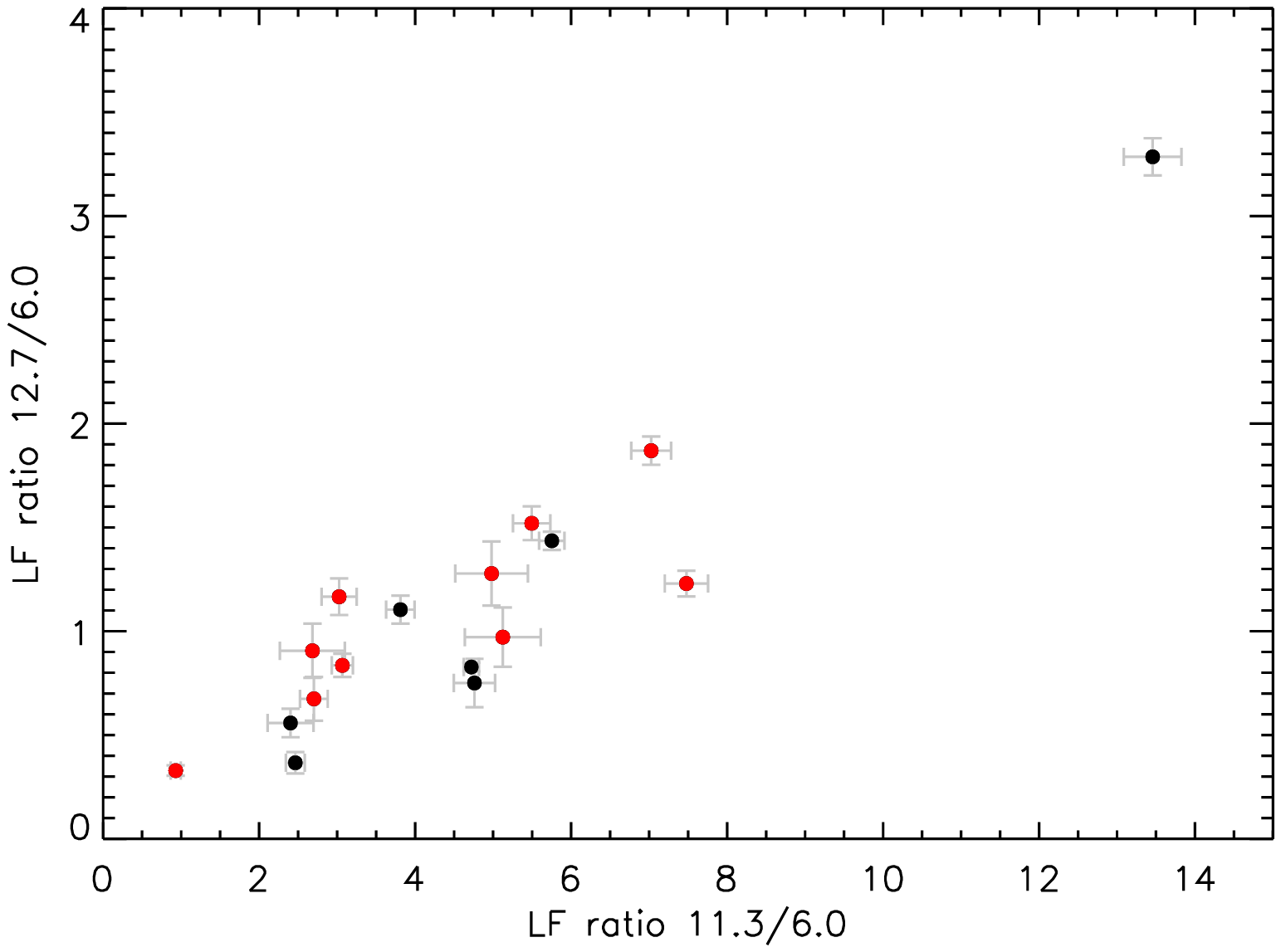}
\caption{Strong correlation between the line flux ratios
  6.2/6.0 and 7.8/6.0 (p-value $1 \times 10^{-6}$, {\em top}) and
  12.7/6.0 and 11.3/6.0 (p-value $2 \times 10^{-4}$, {\em
    bottom}). The 6.2 and 7.8-\mic features are both CC stretching
  modes, the 11.3 and 12.7-\mic features are due to CH out-of-plane
  bending modes.}
           \label{LF6260_7860.ps}%
 \end{figure*}

\clearpage

\begin{figure*}
\centering
\includegraphics[width=\textwidth]{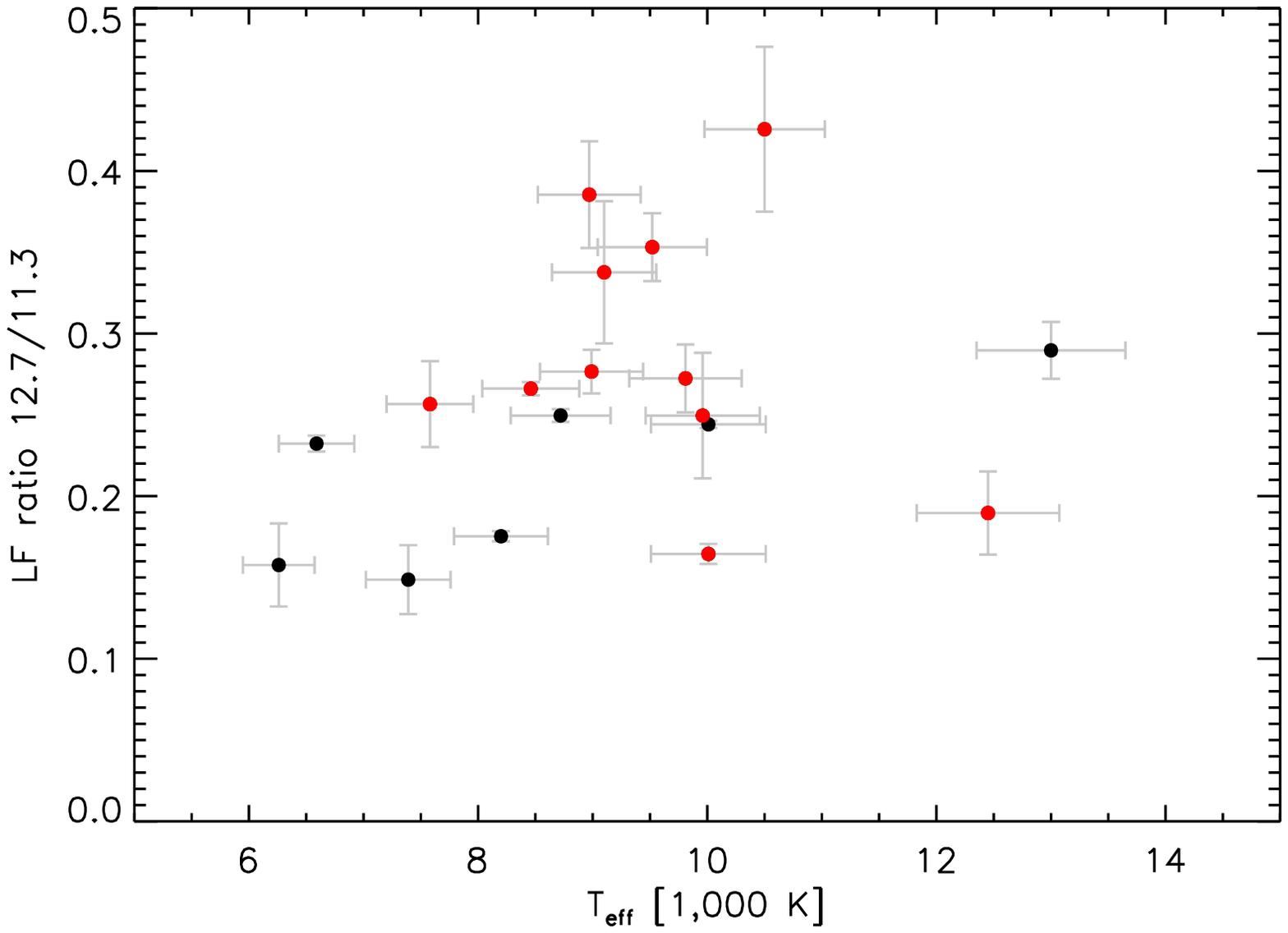}
\caption{Line flux ratio of the CH out-of-plane bending features at
  12.7 (duo+trio CH) and 11.3-\mic (solo CH) as a
  function of stellar effective temperature. A correlation including
  all sample targets can be excluded (p-value 15\%), but is
  tentatively present for the group~I sources alone (p-value 5\%).}
           \label{lfteff_12.7_11.3.ps}%
 \end{figure*}

\begin{figure*}
\centering
\includegraphics[width=\textwidth]{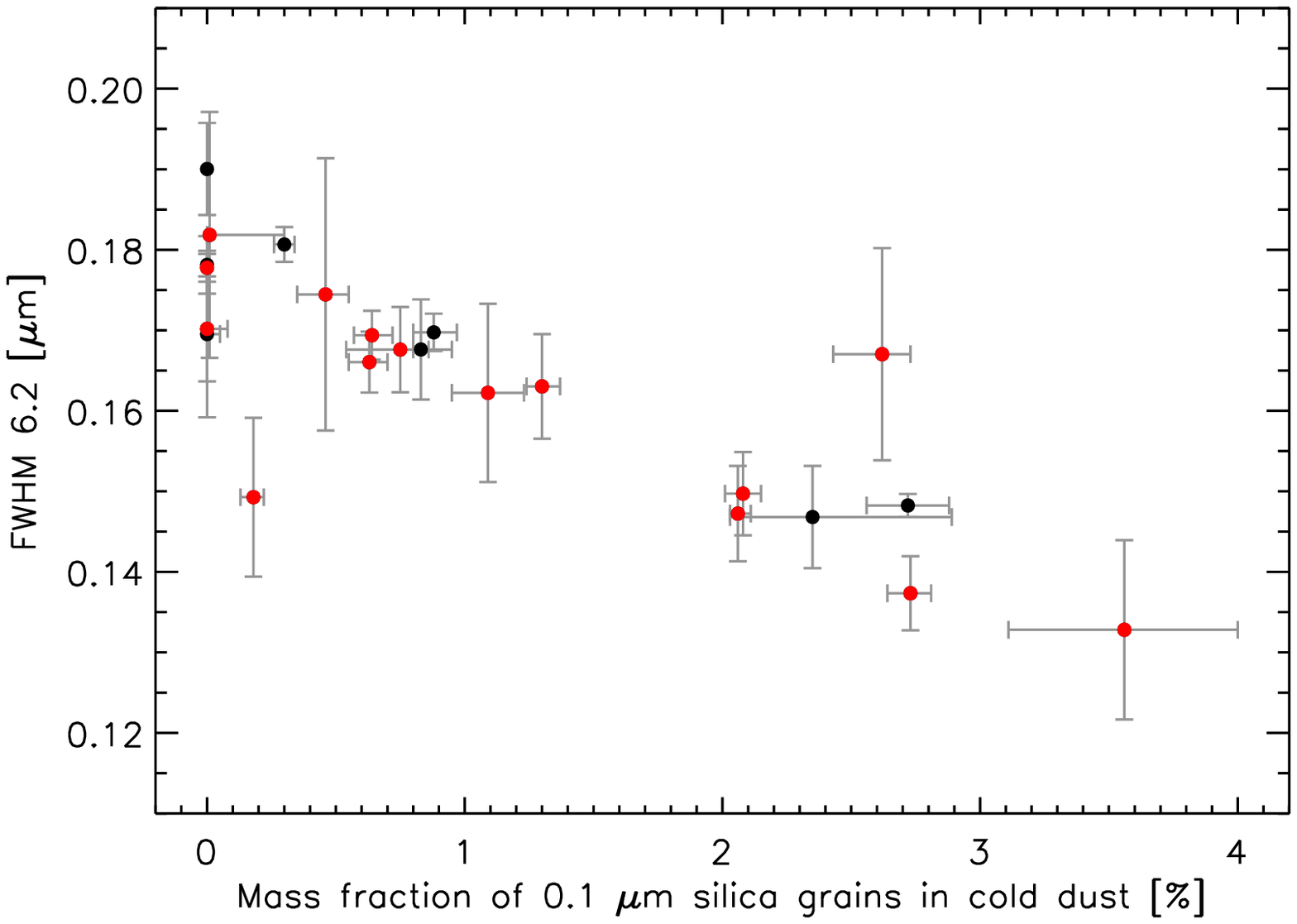}
\caption{Correlation between the FWHM of the PAH 6.2-\mic feature
  versus the mass fraction of the 0.1-\mic silica grains in the cold
  dust component (p-value $1 \times
  10^{-5}$). \label{Silica_FWHM6.2.ps}}
 \end{figure*}

\begin{figure*}
\centering
\includegraphics[width=\textwidth]{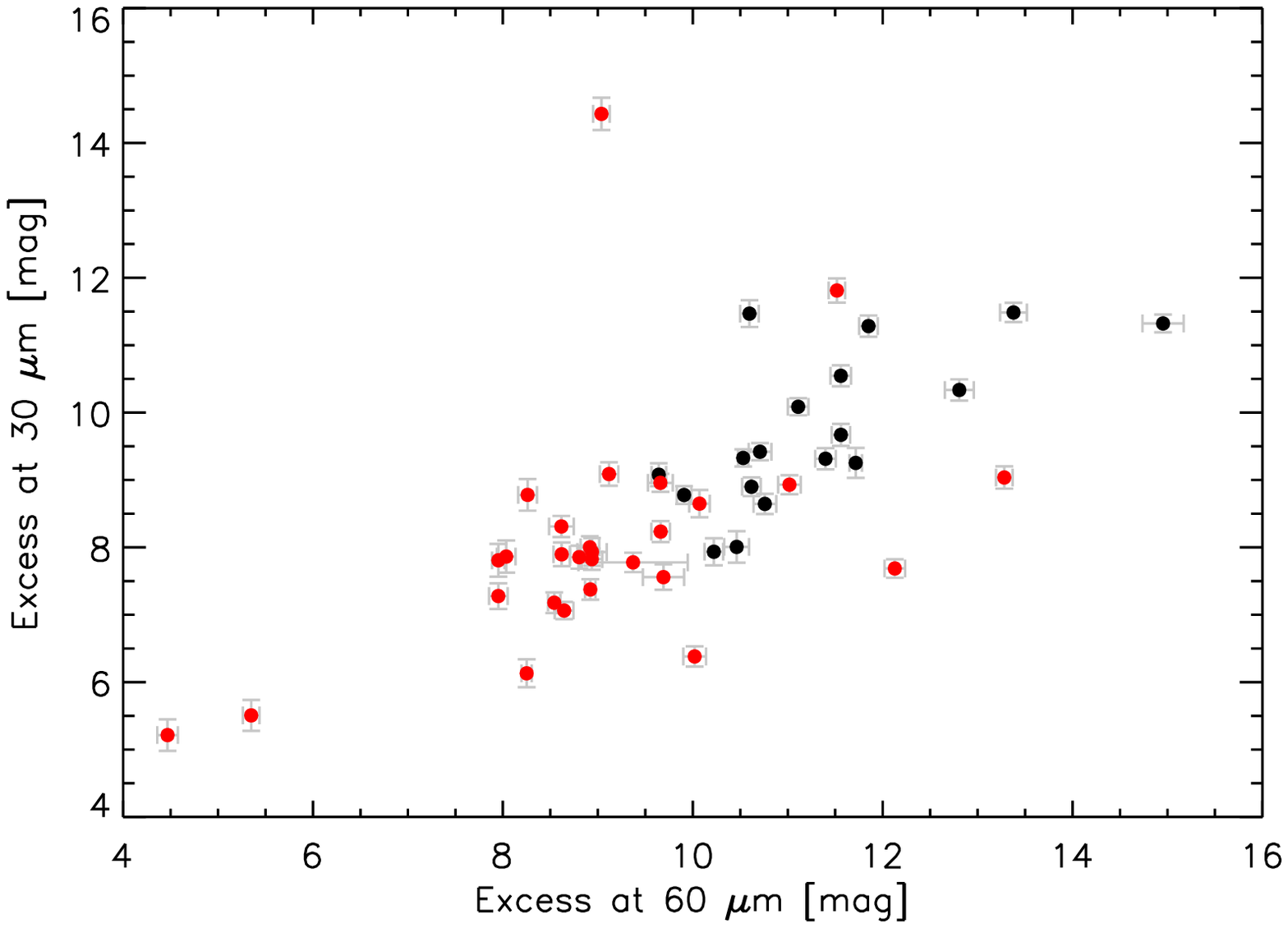}
\caption{The excess magnitude at 30 and 60\,\mic. Both
  are strongly connected (p-value $2 \times 10^{-7}$).}
           \label{exc30vsexc60.ps}%
 \end{figure*}

\begin{figure*}
\centering
\includegraphics[width=\textwidth]{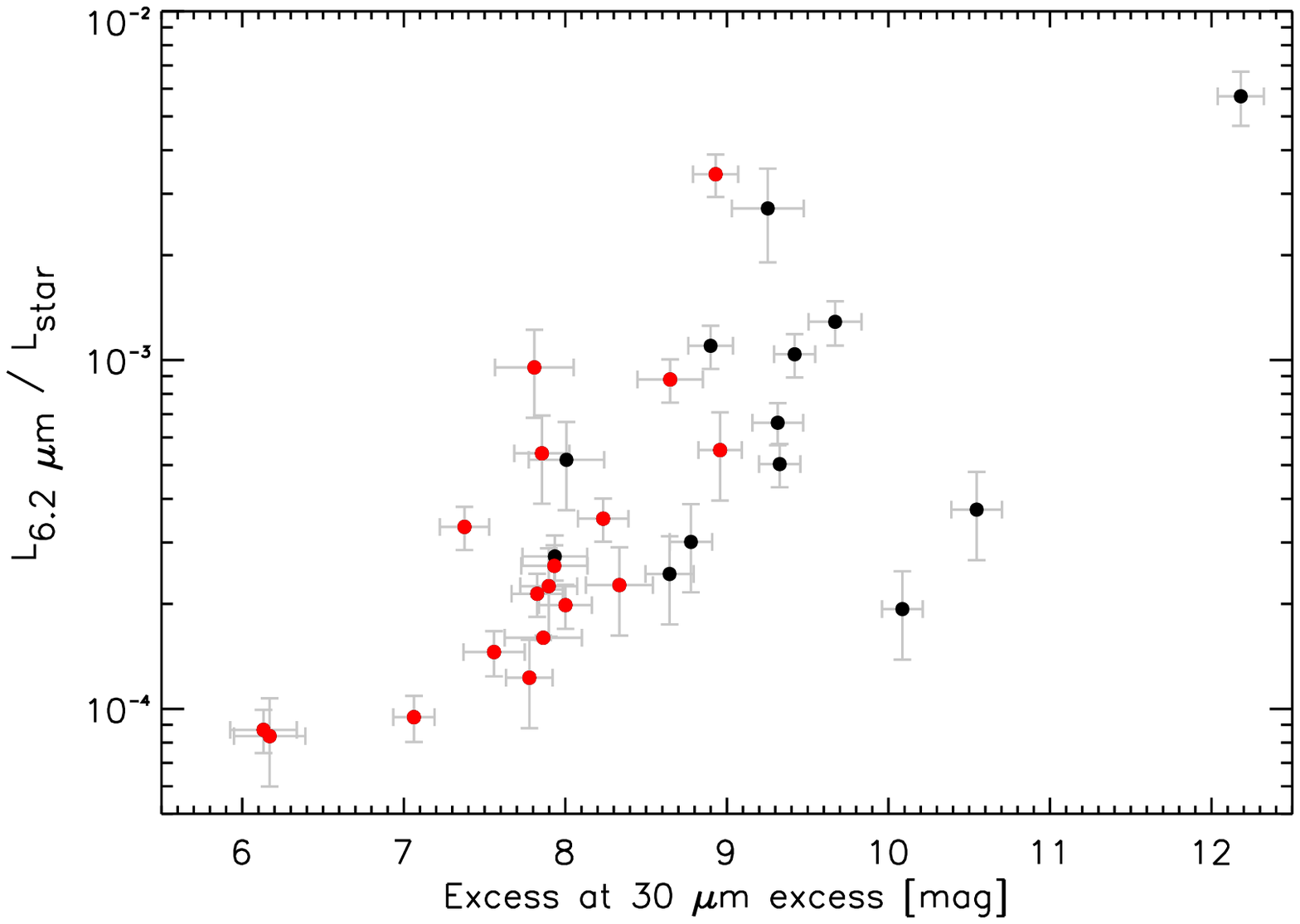}
\caption{The luminosity ratio of the PAH 6.2-\mic CC stretch feature
  and the central star increases with increasing excess at 30-\mic
  (p-value $4 \times 10^{-5}$).}
           \label{PAH62vsEXC30.ps}%
 \end{figure*}

\begin{figure*}
\centering
\includegraphics[width=\textwidth]{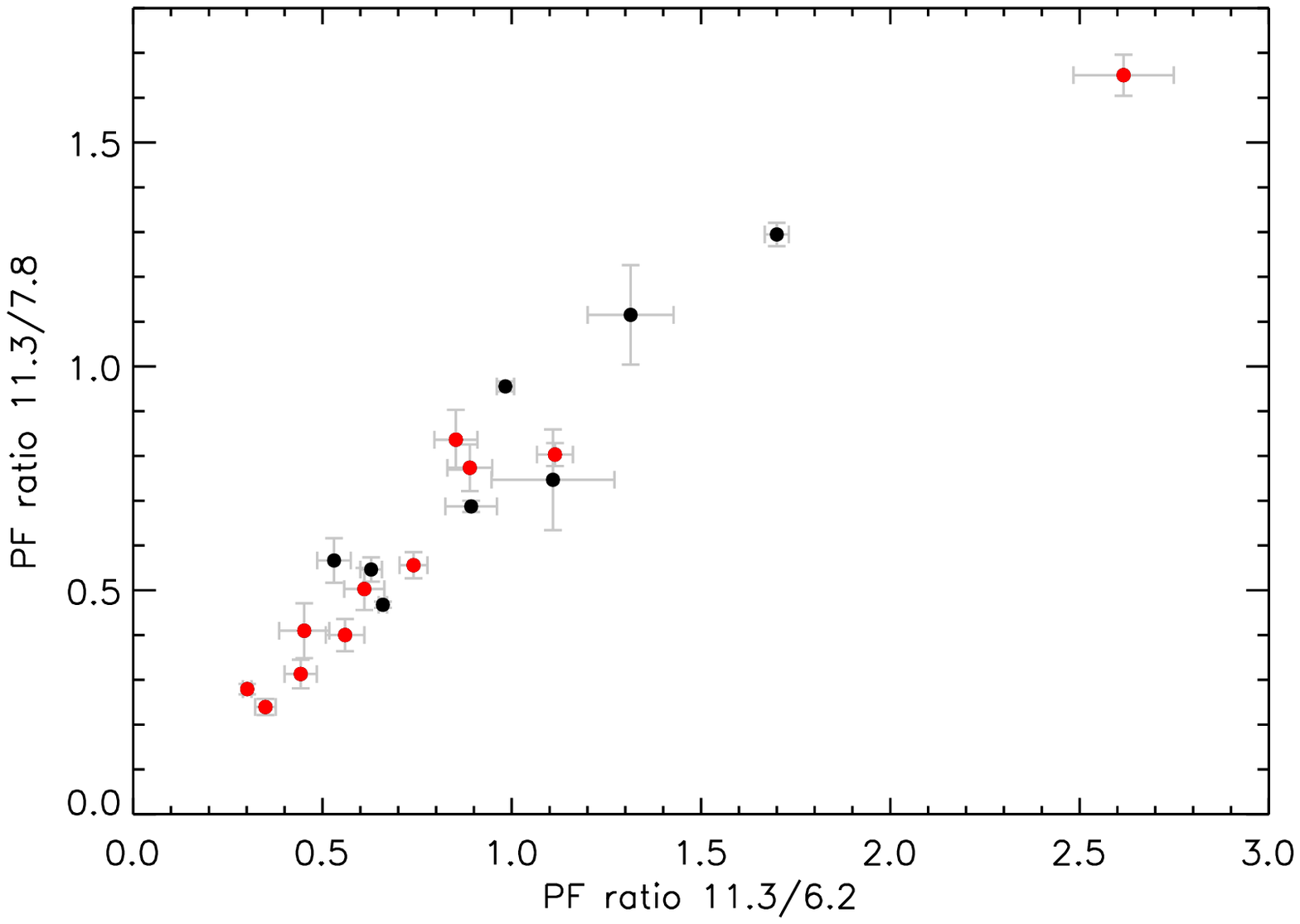}
\caption{Correlation between the peak flux ratios 11.3/6.2 and 11.3/7.8
  (p-value $2 \times 10^{-6}$). Note
that the group~II sources (red dots) have lower values than group~I
sources (black) for both ratios.}
           \label{PF11_62and11_78.ps}%
 \end{figure*}

\clearpage

\begin{figure*}
\centering
\includegraphics[width=\textwidth]{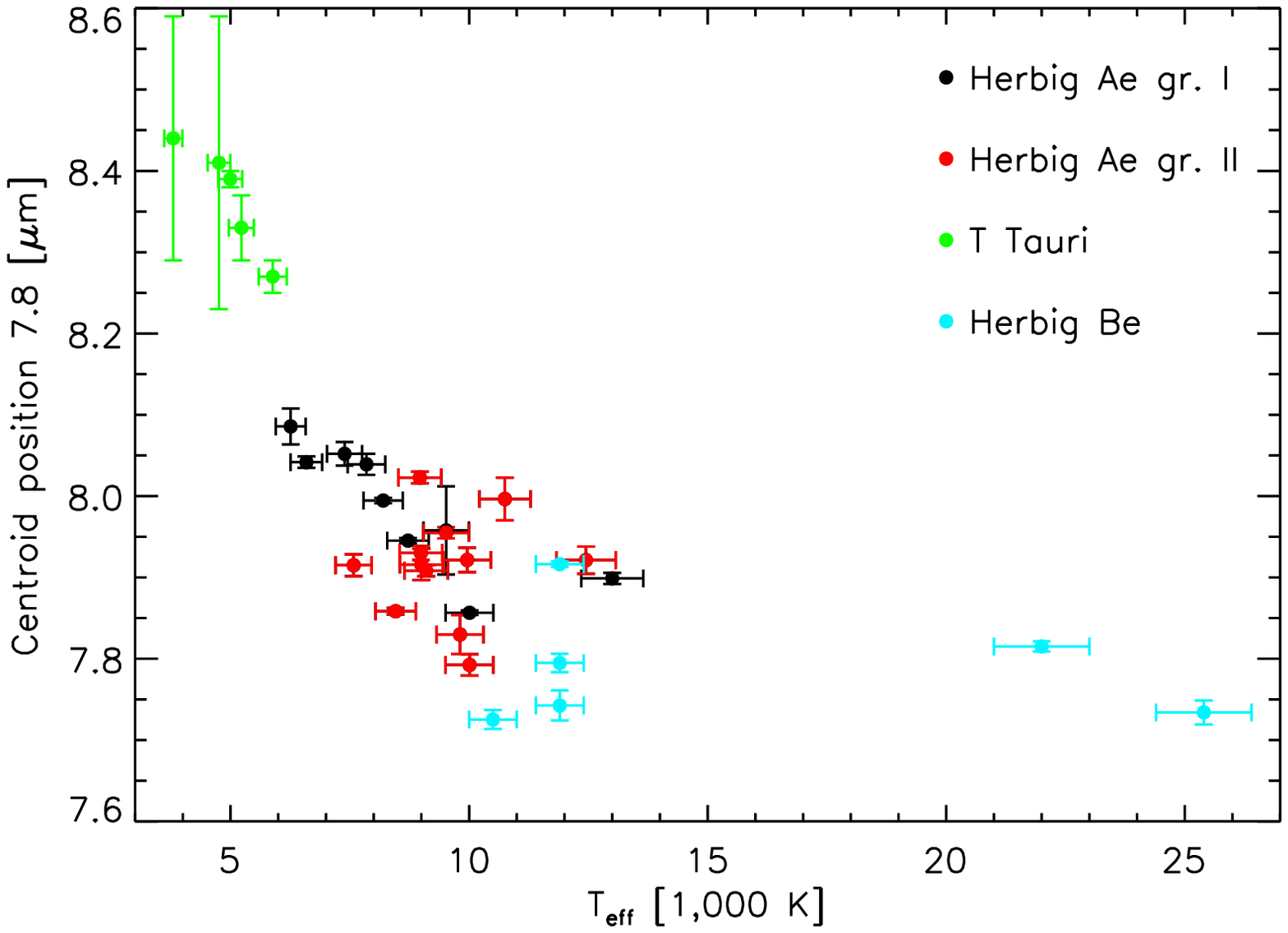}
\caption{The centroid position of the 7.8-\mic feature decreases with
  increasing stellar effective temperature. We have included the
  T~Tauri stars with detected PAH emission from \citet{bouwman08}. The
  Herbig Be stars were observed under Spitzer program PID 50180. The
  correlation is significant for the Herbig Ae 
  stars alone (p-value $4 \times 10^{-3}$) and even more so for the
  entire sample of young stars (p-value $7 \times 10^{-8}$).} 
           \label{centeff_7.8.ps}%
 \end{figure*}

\begin{figure*}
\centering
\includegraphics[width=\textwidth]{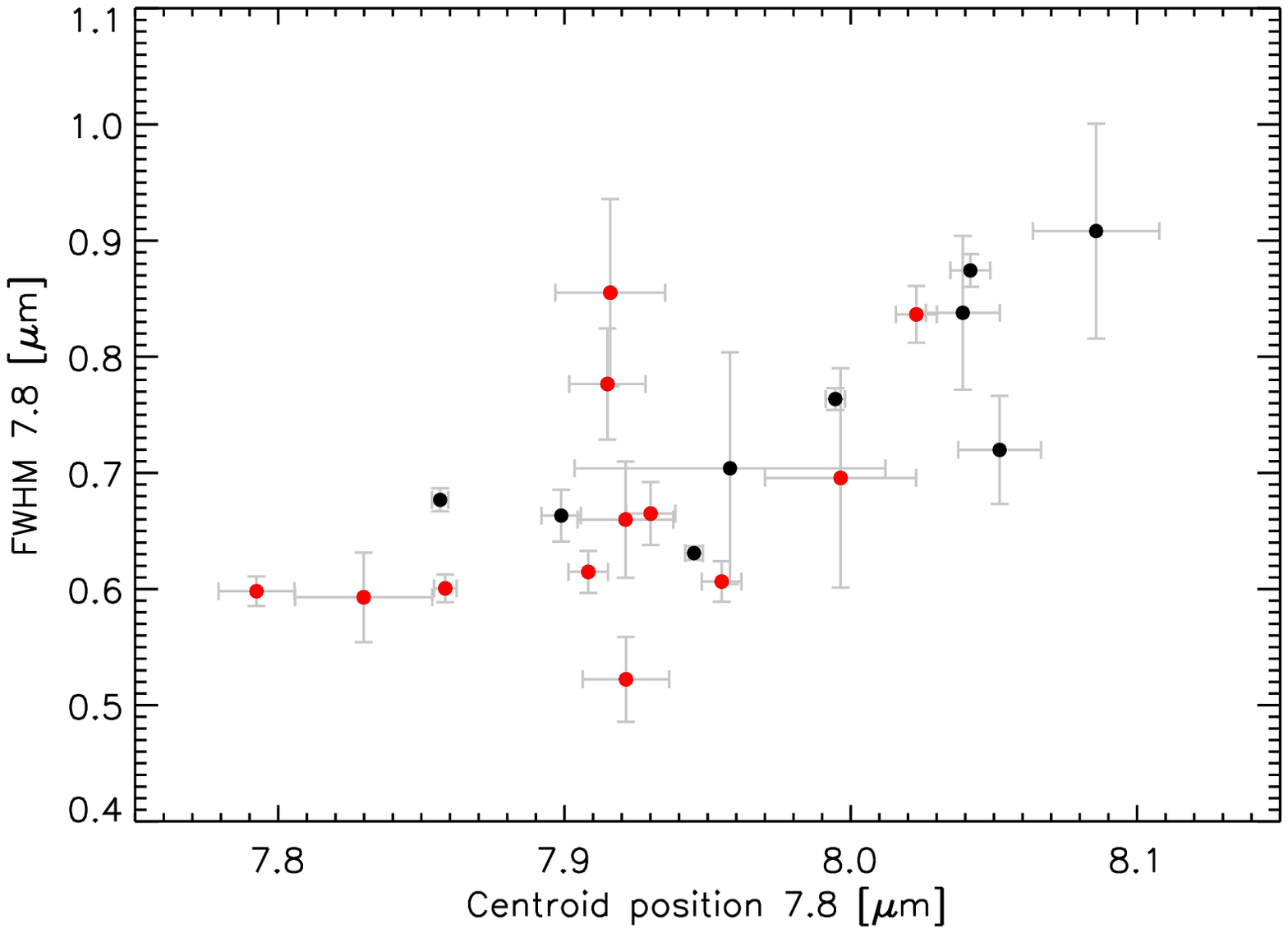}
\caption{A red 7.8-\mic feature has a larger FWHM (p-value $2 \times
  10^{-3}$).}
           \label{cen_vs_fwhm_78.ps}%
 \end{figure*}

\begin{figure*}
\centering
\includegraphics[width=\textwidth]{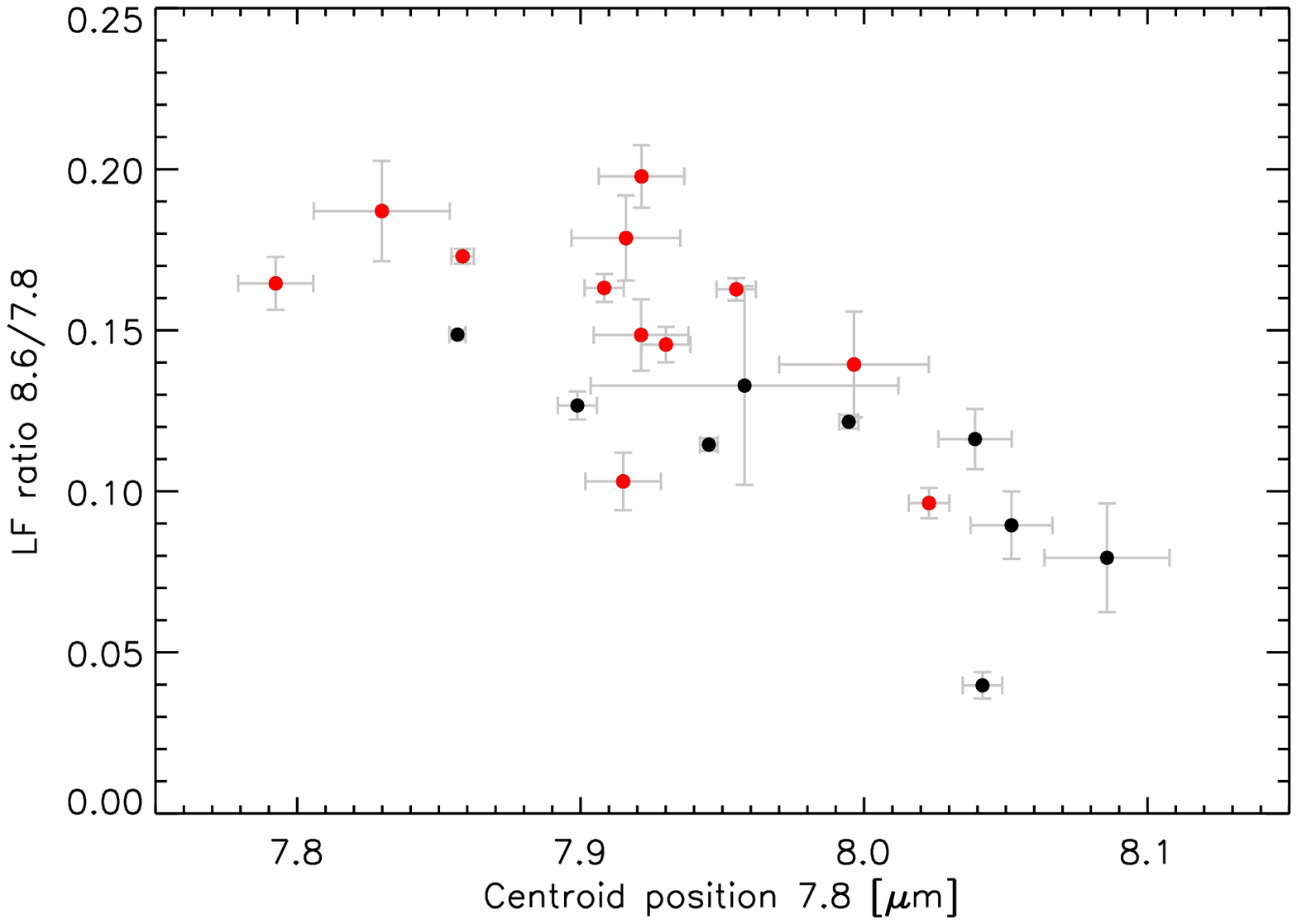}
\caption{When the centroid wavelength of the 7.8-\mic
  feature increases, the 8.6/7.8 line flux ratio decreases (p-value $4
  \times 10^{-4}$). The 7.8-\mic feature is due to CC stretches, the
  8.6-\mic feature is attributed to a CH in-plane bending mode in
  aromatic hydrocarbons.}
           \label{cen78_vs_LF86_78.ps}%
 \end{figure*}

\begin{figure*}
\centering
\includegraphics[width=\textwidth]{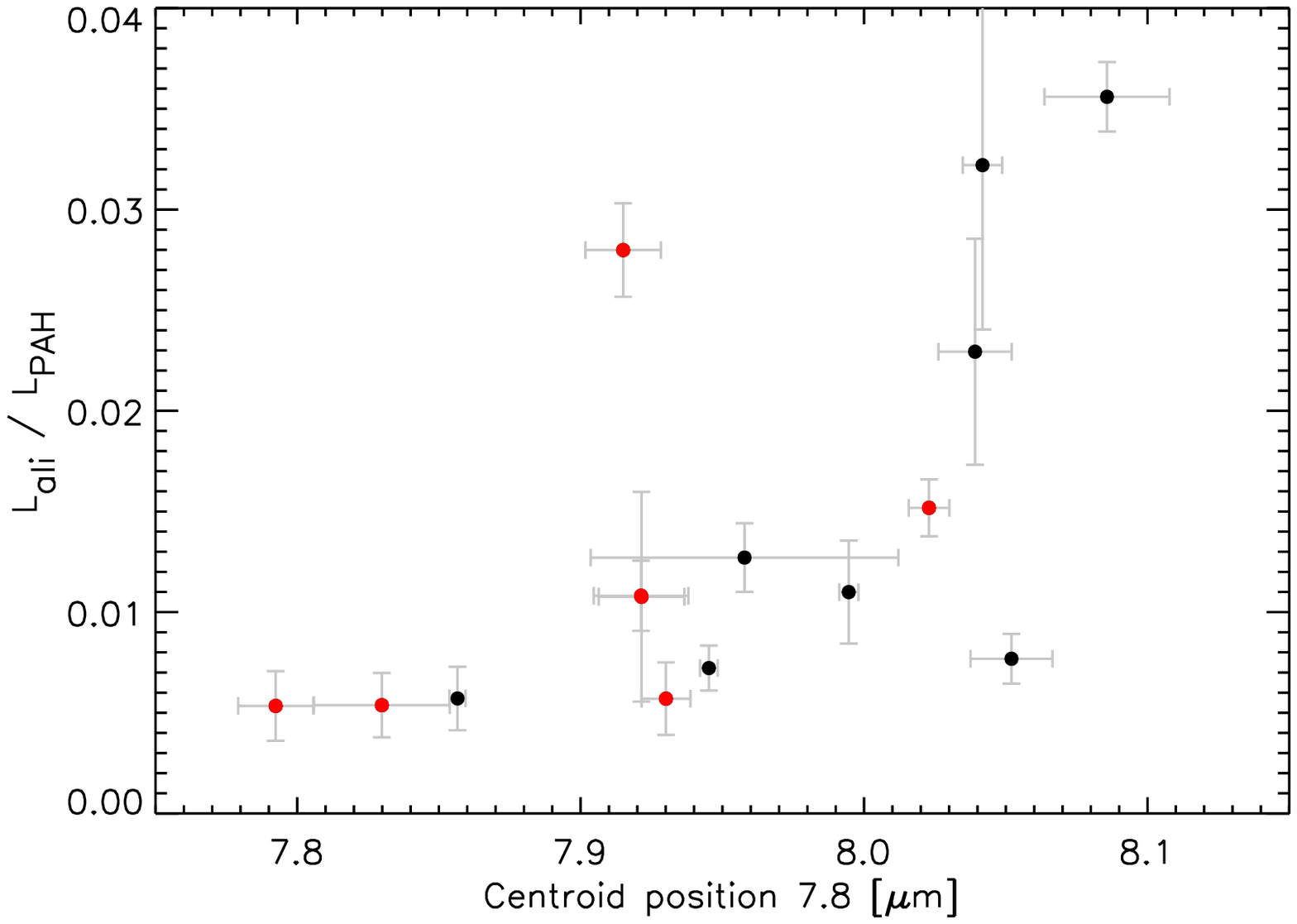}
\caption{Correlation between the aliphatic-to-PAH luminosity ratio and
  the centroid position of the 7.8-\mic feature (p-value $4 \times
  10^{-3}$).}
           \label{LaliLpahvscen78.ps}%
 \end{figure*}

\clearpage


\begin{deluxetable}{lccccccc}
\tabletypesize{\scriptsize}
\tablecaption{Stellar and disk parameters \label{table_sed_parameters}}
\tablewidth{0pt}
\tablehead{
Target       & Group & \Teff & M(7\,\mic)$^a$ & M(30\,\mic)$^a$ & [13.5/7]$^b$
& [30/13.5]$^b$ \\
             &       & (K)   & (mag)          & (mag)           & &    
}
\startdata
HD\,31293       & I & 9520 &  4.3$\pm$0.1 & 9.3$\pm$0.2 &  1.14$\pm$0.05 &  4.5$\pm$0.1 \\
HD\,31648       & II& 8720 &  4.6$\pm$0.1 & 8.2$\pm$0.2 &  1.28$\pm$0.03 & 1.19$\pm$0.03 \\
HD\,34282       & I & 8720 &  4.1$\pm$0.1 & 9.3$\pm$0.2 &  0.56$\pm$0.09 & 10.0$\pm$0.6 \\
HD\,35187       & II& 8970 &  2.8$\pm$0.1 & 7.8$\pm$0.2 &   2.5$\pm$0.2  &  2.1$\pm$0.1 \\
HD\,35929       & II& 6870 &  2.0$\pm$0.1 & 3.9$\pm$0.2 &  0.70$\pm$0.03 & 0.43$\pm$0.01 \\
HD\,36112       & I & 7850 &  3.6$\pm$0.1 & 8.6$\pm$0.1 &  1.37$\pm$0.05 &  4.1$\pm$0.2 \\
HD\,244604      & II& 8730 &  3.8$\pm$0.1 & 7.9$\pm$0.2 &  1.51$\pm$0.05 & 1.43$\pm$0.03 \\
HD\,36917       & II& 10010&  2.8$\pm$0.1 & 6.2$\pm$0.2 &  0.86$\pm$0.06 & 1.35$\pm$0.06 \\
HD\,37258       & II& 8970 &  4.1$\pm$0.1 & 7.9$\pm$0.2 &  1.65$\pm$0.04 & 1.10$\pm$0.03 \\
BF Ori        & II& 8985 &  4.2$\pm$0.1 & 7.6$\pm$0.2 &  1.36$\pm$0.03 & 0.87$\pm$0.03 \\
HD\,37357       & II& 9230 &  3.3$\pm$0.1 & 7.8$\pm$0.1 &  1.70$\pm$0.06 & 1.85$\pm$0.04 \\
HD\,37411       & II& 9100 &  4.1$\pm$0.1 & 9.0$\pm$0.1 &  1.12$\pm$0.08 &  4.0$\pm$0.2 \\
RR Tau        & II& 8460 &  5.0$\pm$0.1 & 8.9$\pm$0.1 &   1.1$\pm$0.1  & 1.72$\pm$0.04 \\
HD\,37806       & II& 9480 &  5.1$\pm$0.1 & 8.3$\pm$0.2 &  1.25$\pm$0.03 & 0.85$\pm$0.07 \\
HD\,38120       & I & 10500&  4.7$\pm$0.1 &10.5$\pm$0.2 &   4.5$\pm$0.5  &  2.6$\pm$0.2 \\
HD\,250550      & I & 10700&  5.0$\pm$0.1 &10.1$\pm$0.1 &   2.2$\pm$0.1  &  2.5$\pm$0.1 \\
HD\,259431      & I & 25400&  --          &10.3$\pm$0.2 &  --            & 2.55$\pm$0.05 \\
V590 Mon      & I & 13000&  6.8$\pm$0.1 &12.2$\pm$0.1 &   3.0$\pm$0.6  &  2.2$\pm$0.1 \\
HD\,50138       & II& 12230&  --          &14.4$\pm$0.2 &  --            & 0.78$\pm$0.04 \\
NX Pup        & II& 7290 &  --          &11.8$\pm$0.2 &  --            & 0.78$\pm$0.02 \\
HD\,58647       & II& 10750&  3.6$\pm$0.1 & 5.5$\pm$0.2 &  0.66$\pm$0.02 & 0.49$\pm$0.01 \\
HD\,72106S      & II& 9810 &  3.4$\pm$0.1 & 7.9$\pm$0.2 &   2.0$\pm$0.2  & 1.62$\pm$0.07 \\
HD\,85567       & II& 12450&  4.9$\pm$0.1 & 7.9$\pm$0.2 &  1.02$\pm$0.02 & 0.82$\pm$0.04 \\
HD\,95881       & II& 8990 &  4.9$\pm$0.1 & 7.8$\pm$0.2 &  1.00$\pm$0.03 & 0.77$\pm$0.05 \\
HD\,97048       & I & 10010&  3.9$\pm$0.1 & 9.7$\pm$0.2 &   1.6$\pm$0.4  &  5.9$\pm$0.4 \\
HD\,98922       & II& 10500&  --          & 8.8$\pm$0.2 &  --            & 0.75$\pm$0.03 \\
HD\,100453      & I & 7390 &  3.9$\pm$0.1 & 9.4$\pm$0.1 &   1.7$\pm$0.1  &  5.2$\pm$0.3 \\
HD\,100546      & I & 10500&  --          &11.3$\pm$0.2 &  --            &  3.5$\pm$0.2 \\
HD\,101412      & II& 9960 &  4.6$\pm$0.1 & 7.9$\pm$0.2 &  1.31$\pm$0.04 & 0.92$\pm$0.03 \\
HD\,104237      & II& 8405 &  3.4$\pm$0.1 & 7.2$\pm$0.2 &  1.31$\pm$0.02 & 1.28$\pm$0.03 \\
SS73 44       & I & 15000&  6.6$\pm$0.1 &11.5$\pm$0.1 &  2.10$\pm$0.09 & 2.40$\pm$0.08 \\
HD\,135344B     & I & 6590 &  3.4$\pm$0.1 & 7.9$\pm$0.2 &  0.32$\pm$0.02 & 10.9$\pm$0.3 \\
HD\,139614      & I & 7850 &  3.0$\pm$0.1 & 9.3$\pm$0.1 &   4.1$\pm$0.3  &  4.2$\pm$0.3 \\
HD\,141569      & II& 9520 &  1.1$\pm$0.1 & 6.1$\pm$0.2 &  0.73$\pm$0.07 &  6.8$\pm$0.2 \\
HD\,142666      & II& 7580 &  3.3$\pm$0.1 & 7.4$\pm$0.2 &  1.45$\pm$0.06 & 1.53$\pm$0.05 \\
HD\,142527      & I & 6260 &  3.1$\pm$0.1 & 8.0$\pm$0.2 &  0.95$\pm$0.02 &  5.0$\pm$0.1 \\
HD\,144432      & II& 7345 &  3.5$\pm$0.1 & 8.0$\pm$0.2 &  1.82$\pm$0.06 & 1.82$\pm$0.06 \\
HD\,144668      & II& 7930 &  --          & 8.3$\pm$0.2 &  --            & 0.96$\pm$0.02 \\
HD\,145263      & II& 7200 &  0.8$\pm$0.2 & 5.4$\pm$0.2 &   1.9$\pm$0.1  &  2.0$\pm$0.1 \\
Wray 15-1484  & II& 30000&  --          & 9.0$\pm$0.2 &  --            & 0.76$\pm$0.02 \\
HD\,150193      & II& 8990 &  --          & 9.1$\pm$0.2 &  --            & 1.42$\pm$0.05 \\
HD\,152404      & II& 6450 &  2.5$\pm$0.1 & 7.1$\pm$0.1 &  1.13$\pm$0.05 &  3.3$\pm$0.1 \\
KK Oph        & II& 8030 &  --          & 7.7$\pm$0.1 &  --            & 1.04$\pm$0.02 \\
HD\,158643      & II& 10100&  2.7$\pm$0.1 & 5.2$\pm$0.2 &  0.96$\pm$0.02 & 0.59$\pm$0.02 \\
HD\,163296      & II& 8720 &  --          & 6.4$\pm$0.2 &  --            &  2.0$\pm$0.1 \\
HD\,169142      & I & 8200 &  2.8$\pm$0.1 & 8.9$\pm$0.1 &   1.7$\pm$0.3  &  7.8$\pm$0.5 \\
VV Ser        & II& 9000 &  5.7$\pm$0.1 & 8.6$\pm$0.2 &  1.09$\pm$0.02 & 0.79$\pm$0.02 \\
T CrA         & I & 7200 &  --          &11.3$\pm$0.1 &  --            &  5.0$\pm$0.3 \\
HD\,179218      & I & 9810 &  --          &11.5$\pm$0.2 &  --            &  2.4$\pm$0.2 \\
WW Vul        & I & 8430 &  4.4$\pm$0.1 & 8.8$\pm$0.1 &  1.80$\pm$0.05 & 1.60$\pm$0.04 \\
HD\,190073      & II& 8990 &  4.3$\pm$0.1 & 7.3$\pm$0.2 &  1.14$\pm$0.02 & 0.75$\pm$0.02 \\
LkHa 224      & I & 7850 &  --          & 8.9$\pm$0.1 &  --            & 1.99$\pm$0.02 \\
HD\,203024      & I & 8200 &  3.2$\pm$0.1 & 9.1$\pm$0.2 &   5.0$\pm$0.6  &  2.3$\pm$0.2 \\
\enddata
\tablecomments{(a) The excess magnitude M($\lambda$) is relative to
  the stellar photosphere at $\lambda$; (b) Continuum flux ratio at
  the given wavelengths.}
\end{deluxetable}

\begin{deluxetable}{lllccccccc}
\tabletypesize{\scriptsize}
\rotate
\tablecaption{Sample-averaged parameters of the detected features\label{sample_avg}}
\tablewidth{0pt}
\tablehead{
Feature & Limits & Mode & Ref. & Det. & $<$PP$>$  & $<$Cen$>$ &
$<$FWHM$>$ &  
$<$$\log_{10}\mathrm{L}_f/\mathrm{L}_\star$$>$  &
$<$$\log_{10}\mathrm{L}_f/\mathrm{L}_\mathrm{PAH}$$>$  \\
        & (\mic) & & & (\%) & (\mic) & (\mic) & (\mic) & 
}
\startdata
5.7    & 5.58$-$5.83   & CH-oop?           & 1   & 45 &  5.69  $\pm$ 0.02   &    5.70 $\pm$  0.02 &   0.07 $\pm$  0.02  &  $-$4.8 $\pm$ 0.6 & $-$2.2  $\pm$ 0.3  \\
6.0    & 5.83$-$6.10   & carbonyl-s?       & 1   & 78 &  6.00  $\pm$ 0.04   &    6.00 $\pm$  0.03 &   0.11 $\pm$  0.02  &  $-$4.2 $\pm$ 0.4 & $-$1.5  $\pm$ 0.3  \\
6.2    & 5.83$-$6.66   & CC-s (PANH?)      & 1,5 & 78 &  6.26  $\pm$ 0.02   &    6.28 $\pm$  0.02 &   0.16 $\pm$  0.01  &  $-$3.4 $\pm$ 0.5 & $-$0.7  $\pm$ 0.2  \\
6.8    & 6.66$-$7.06   & CH$_2$ and CH$_3$ (ali) & 2   & 38 &  6.87  $\pm$ 0.03   &    6.85 $\pm$  0.07 &   0.11 $\pm$  0.04  &  $-$4.7 $\pm$ 0.3 & $-$2.0  $\pm$ 0.3  \\
       &               & or CH-ip n        & 5   &    &                     &                     &                     &                   &                    \\
7.2    & 7.06$-$7.36   & CH$_3$ (ali)      & 2   & 50 &  7.22  $\pm$ 0.02   &    7.22 $\pm$  0.01 &   0.11 $\pm$  0.03  &  $-$4.6 $\pm$ 0.2 & $-$2.0  $\pm$ 0.3  \\
7.8    & 7.06$-$9.10   & CC-s              & 1,5 & 53 &  7.88  $\pm$ 0.08   &    7.94 $\pm$  0.08 &   0.7  $\pm$   0.1  &  $-$2.8 $\pm$ 0.6 & $-$0.24 $\pm$ 0.06 \\
8.6    & 8.25$-$9.10   & CH-ip             & 1,5 & 50 &  8.66  $\pm$ 0.03   &    8.64 $\pm$  0.03 &   0.25 $\pm$  0.04  &  $-$3.7 $\pm$ 0.6 & $-$1.1  $\pm$ 0.2  \\
10.6   & 10.35$-$10.90 & CH-oop s +        & 3   & 26 & 10.64  $\pm$ 0.04   &   10.61 $\pm$  0.06 &   0.25 $\pm$  0.09  &  $-$4.5 $\pm$ 0.7 & $-$2.1  $\pm$ 0.3  \\
11.3   & 10.35$-$12.30 & CH-oop s n($-$)   & 3,5 & 42 & 11.27  $\pm$ 0.02   &   11.3  $\pm$  0.1  &   0.29 $\pm$  0.05  &  $-$3.6 $\pm$ 0.5 & $-$1.0  $\pm$ 0.2  \\
12.0   & 11.70$-$12.30 & CH-oop d n+$-$    & 3   & 28 & 12.03  $\pm$ 0.05   &   12.00 $\pm$  0.03 &   0.22 $\pm$  0.08  &  $-$4.5 $\pm$ 0.6 & $-$1.9  $\pm$ 0.3  \\
12.7   & 12.10$-$13.10 & CH-oop dt n+$-$   & 3,5 & 38 & 12.7   $\pm$ 0.1    &   12.61 $\pm$  0.05 &   0.41 $\pm$  0.09  &  $-$4.2 $\pm$ 0.6 & $-$1.7  $\pm$ 0.1  \\
13.5   & 13.20$-$13.90 & CH-oop q n+$-$    & 3,5 & 19 & 13.59  $\pm$ 0.04   &   13.6  $\pm$  0.1  &   0.19 $\pm$  0.07  &  $-$5.2 $\pm$ 0.5 & $-$2.7  $\pm$ 0.4  \\
16$-$19& 15.60$-$18.10 & CCC-oop           & 4   &  9 & 16.7   $\pm$ 0.4    &   16.8  $\pm$  0.3  &   0.8  $\pm$   0.3  &  $-$4.0 $\pm$ 0.5 & $-$1.6  $\pm$ 0.4 \\
\enddata
\tablecomments{ The
  error bar represents the standard deviation in the sample. Det. =
  detection rate; PP =
  Peak position; Cen = Centroid position; FWHM = Full width at half
  maximum; $\mathrm{L}_f / \mathrm{L}_\star$ = Feature-to-stellar
  luminosity ratio; $\mathrm{L}_f/\mathrm{L}_\mathrm{PAH}$ = 
  Feature-to-total-PAH luminosity ratio. The
  integration limits of the features are given in the second
  column. The third column summarizes the identification 
  of the mode of the feature, as published in the literature. 
  CH = carbon-hydrogen bond; CC = carbon-carbon bond. oop = out-of-plane 
  bending mode; ip = in-plane bending or wagging mode; s = stretching
  mode. PANH = nitrogenated PAH; ali = aliphatic.
  For the features in the 10 -- 14-\mic region:
  s = solo or non-adjacent CH group; d = duo; t = trio;
  q = quartet CH group; n = neutral PAHs; + = cationic PAHs; $-$ =
  anionic PAHS.}
\tablerefs{(1) \citeauthor{peeters02} \citeyear{peeters02};
  (2) \citeauthor{dartois07} \citeyear{dartois07}; (3)
  \citeauthor{hony01} \citeyear{hony01}; (4) \citeauthor{peeters04}
  \citeyear{peeters04}; (5) \citeauthor{bauschlicher09}
  \citeyear{bauschlicher09}.}
\end{deluxetable}


\begin{deluxetable}{lccccccc}
\tabletypesize{\scriptsize}
\tablecaption{Line fluxes of the PAH and aliphatic features in the
  5--10-\mic range \label{table_6_lineflux}}
\tablewidth{0pt}
\tablehead{
Target       &5.7 &6.0 &6.2 &6.8 &7.2 & 7.8 & 8.6
}
\startdata
HD\,31293      & $<$5 ($-$13)           & 72$\pm$1 ($-$13)     & 475$\pm$4 ($-$13)    & 24$\pm$4 ($-$13)     & 15$\pm$2 ($-$13)     & 181$\pm$6 ($-$12)    & 24$\pm$6 ($-$12)     \\
HD\,31648      & $<$7 ($-$13)           & 224$\pm$9 ($-$14)    & 90$\pm$1 ($-$13)     & 8$\pm$3 ($-$13)      & 92$\pm$8 ($-$14)     & $<$2 ($-$11)           & $<$5 ($-$12)           \\
HD\,34282      & 3$\pm$1 ($-$13)      & 96$\pm$3 ($-$14)     & 1010$\pm$5 ($-$14)   & 17$\pm$4 ($-$14)     & 25$\pm$5 ($-$14)     & 3380$\pm$8 ($-$14)   & 387$\pm$6 ($-$14)    \\
HD\,35187      & 7$\pm$4 ($-$14)      & 88$\pm$4 ($-$14)     & 839$\pm$6 ($-$14)    & 35$\pm$5 ($-$14)     & 37$\pm$5 ($-$14)     & 313$\pm$2 ($-$13)    & 30$\pm$1 ($-$13)     \\
HD\,35929      & $<$1 ($-$13)           & $<$7 ($-$14)           & $<$1 ($-$13)           & $<$4 ($-$13)           & $<$1 ($-$13)           & $<$5 ($-$13)           & $<$2 ($-$13)           \\
HD\,36112      & 12$\pm$9 ($-$14)     & 41$\pm$6 ($-$14)     & 414$\pm$7 ($-$14)    & 36$\pm$7 ($-$14)     & 3$\pm$1 ($-$13)      & $<$1 ($-$11)           & $<$1 ($-$12)           \\
HD\,244604     & $<$2 ($-$13)           & 37$\pm$3 ($-$14)     & 148$\pm$3 ($-$14)    & $<$1 ($-$13)           & $<$8 ($-$14)           & $<$2 ($-$12)           & $<$9 ($-$13)           \\
HD\,36917      & 8$\pm$3 ($-$14)      & 71$\pm$3 ($-$14)     & 364$\pm$4 ($-$14)    & 14$\pm$5 ($-$14)     & $<$4 ($-$13)           & 123$\pm$1 ($-$13)    & 20$\pm$1 ($-$13)     \\
HD\,37258      & $<$2 ($-$13)           & $<$9 ($-$14)           & $<$1 ($-$13)           & $<$2 ($-$13)           & $<$9 ($-$14)           & $<$7 ($-$13)           & $<$5 ($-$13)           \\
BF Ori       & $<$9 ($-$14)           & 11$\pm$1 ($-$14)     & 61$\pm$3 ($-$14)     & $<$6 ($-$14)           & $<$8 ($-$14)           & $<$1 ($-$12)           & $<$4 ($-$13)           \\
HD\,37357      & $<$9 ($-$14)           & 24$\pm$2 ($-$14)     & 129$\pm$3 ($-$14)    & 9$\pm$5 ($-$14)      & $<$2 ($-$13)           & $<$3 ($-$12)           & $<$6 ($-$13)           \\
HD\,37411      & 11$\pm$2 ($-$14)     & 24$\pm$3 ($-$14)     & 235$\pm$4 ($-$14)    & $<$9 ($-$14)           & $<$7 ($-$14)           & 912$\pm$6 ($-$14)    & 149$\pm$4 ($-$14)    \\
RR Tau       & 48$\pm$7 ($-$14)     & 94$\pm$3 ($-$14)     & 922$\pm$8 ($-$14)    & $<$1 ($-$13)           & $<$2 ($-$13)           & 311$\pm$1 ($-$13)    & 537$\pm$7 ($-$14)    \\
HD\,37806      & $<$5 ($-$13)           & 89$\pm$6 ($-$14)     & 479$\pm$9 ($-$14)    & $<$4 ($-$13)           & 81$\pm$8 ($-$14)     & $<$1 ($-$11)           & $<$3 ($-$12)           \\
HD\,38120      & $<$1 ($-$13)           & 59$\pm$6 ($-$14)     & 412$\pm$8 ($-$14)    & 19$\pm$7 ($-$14)     & 29$\pm$7 ($-$14)     & $<$7 ($-$12)           & $<$1 ($-$12)           \\
HD\,250550     & 9$\pm$3 ($-$14)      & 37$\pm$3 ($-$14)     & 166$\pm$3 ($-$14)    & 15$\pm$8 ($-$14)     & 21$\pm$2 ($-$14)     & $<$6 ($-$12)           & $<$3 ($-$13)           \\
HD\,259431     &  n.s.                &  n.s.                &  n.s.                &  n.s.                &  n.s.                &  n.s.                &  n.s.                \\
V590 Mon     & $<$2 ($-$13)           & 76$\pm$3 ($-$14)     & 681$\pm$4 ($-$14)    & $<$2 ($-$13)           & $<$3 ($-$13)           & 243$\pm$2 ($-$13)    & 31$\pm$1 ($-$13)     \\
HD\,50138      &  n.s.                &  n.s.                &  n.s.                &  n.s.                &  n.s.                &  n.s.                & $<$8 ($-$12)           \\
NX Pup       &  n.s.                &  n.s.                &  n.s.                &  n.s.                &  n.s.                &  n.s.                &  n.s.                \\
HD\,58647      & $<$9 ($-$13)           & $<$4 ($-$13)           & $<$5 ($-$13)           & $<$6 ($-$13)           & $<$6 ($-$13)           & 83$\pm$2 ($-$13)     & 12$\pm$1 ($-$13)     \\
HD\,72106S     & 29$\pm$5 ($-$14)     & 144$\pm$3 ($-$14)    & 759$\pm$3 ($-$14)    & $<$2 ($-$13)           & 19$\pm$6 ($-$14)     & 162$\pm$3 ($-$13)    & 30$\pm$2 ($-$13)     \\
HD\,85567      & 2$\pm$1 ($-$13)      & 76$\pm$6 ($-$14)     & 73$\pm$1 ($-$13)     & $<$5 ($-$13)           & 43$\pm$7 ($-$14)     & 232$\pm$4 ($-$13)    & 34$\pm$3 ($-$13)     \\
HD\,95881      & 7$\pm$1 ($-$13)      & 163$\pm$6 ($-$14)    & 170$\pm$1 ($-$13)    & $<$6 ($-$13)           & 6$\pm$2 ($-$13)      & 569$\pm$5 ($-$13)    & 83$\pm$3 ($-$13)     \\
HD\,97048      & 30$\pm$4 ($-$13)     & 47$\pm$1 ($-$13)     & 551$\pm$3 ($-$13)    & 15$\pm$3 ($-$13)     & 7$\pm$5 ($-$13)      & 2001$\pm$7 ($-$13)   & 297$\pm$3 ($-$13)    \\
HD\,98922      &  n.s.                &  n.s.                &  n.s.                &  n.s.                &  n.s.                &  n.s.                & 21$\pm$1 ($-$12)     \\
HD\,100453     & 11$\pm$1 ($-$13)     & 39$\pm$1 ($-$13)     & 214$\pm$2 ($-$13)    & $<$9 ($-$13)           & 7$\pm$1 ($-$13)      & 522$\pm$7 ($-$13)    & 47$\pm$5 ($-$13)     \\
HD\,100546     &  n.s.                &  n.s.                &  n.s.                &  n.s.                &  n.s.                &  n.s.                & 23$\pm$2 ($-$12)     \\
HD\,101412     & $<$2 ($-$13)           & 56$\pm$3 ($-$14)     & 301$\pm$6 ($-$14)    & $<$5 ($-$13)           & 2$\pm$1 ($-$13)      & 137$\pm$2 ($-$13)    & 27$\pm$1 ($-$13)     \\
HD\,104237     & $<$6 ($-$13)           & $<$3 ($-$13)           & $<$4 ($-$13)           & $<$7 ($-$13)           & $<$4 ($-$13)           & $<$8 ($-$12)           & $<$6 ($-$12)           \\
SS73 44      & $<$1 ($-$13)           & $<$1 ($-$13)           & $<$1 ($-$13)           & $<$1 ($-$13)           & $<$9 ($-$14)           & $<$8 ($-$13)           & $<$2 ($-$13)           \\
HD\,135344B    & 2$\pm$1 ($-$13)      & 11$\pm$1 ($-$13)     & 37$\pm$2 ($-$13)     & 4$\pm$2 ($-$13)      & 33$\pm$7 ($-$14)     & 1433$\pm$9 ($-$14)   & 57$\pm$6 ($-$14)     \\
HD\,139614     & $<$5 ($-$13)           & 44$\pm$4 ($-$14)     & 712$\pm$5 ($-$14)    & 4$\pm$2 ($-$13)      & 45$\pm$8 ($-$14)     & 212$\pm$3 ($-$13)    & 25$\pm$2 ($-$13)     \\
HD\,141569     & 11$\pm$5 ($-$14)     & 89$\pm$5 ($-$14)     & 524$\pm$6 ($-$14)    & $<$8 ($-$14)           & $<$1 ($-$13)           & 1172$\pm$8 ($-$14)   & 191$\pm$4 ($-$14)    \\
HD\,142666     & 20$\pm$8 ($-$14)     & 84$\pm$7 ($-$14)     & 73$\pm$1 ($-$13)     & 73$\pm$7 ($-$14)     & 43$\pm$6 ($-$14)     & 245$\pm$3 ($-$13)    & 25$\pm$2 ($-$13)     \\
HD\,142527     & 8$\pm$2 ($-$13)      & 339$\pm$8 ($-$14)    & 215$\pm$1 ($-$13)    & 28$\pm$2 ($-$13)     & 14$\pm$1 ($-$13)     & 66$\pm$1 ($-$12)     & 5$\pm$1 ($-$12)      \\
HD\,144432     & 16$\pm$7 ($-$14)     & 47$\pm$5 ($-$14)     & 306$\pm$9 ($-$14)    & 3$\pm$1 ($-$13)      & 42$\pm$5 ($-$14)     & $<$3 ($-$12)           & $<$2 ($-$12)           \\
HD\,144668     &  n.s.                &  n.s.                &  n.s.                &  n.s.                &  n.s.                &  n.s.                & $<$2 ($-$12)           \\
HD\,145263     & $<$2 ($-$13)           & $<$6 ($-$14)           & $<$1 ($-$13)           & $<$7 ($-$14)           & $<$1 ($-$13)           & $<$1 ($-$12)           & $<$2 ($-$13)           \\
Wray 15-1484 &  n.s.                &  n.s.                &  n.s.                &  n.s.                &  n.s.                &  n.s.                &  n.s.                \\
HD\,150193     &  n.s.                &  n.s.                &  n.s.                &  n.s.                &  n.s.                &  n.s.                & $<$9 ($-$12)           \\
HD\,152404     & $<$2 ($-$13)           & 15$\pm$4 ($-$14)     & 129$\pm$5 ($-$14)    & $<$2 ($-$13)           & $<$2 ($-$13)           & $<$1 ($-$12)           & $<$4 ($-$13)           \\
KK Oph       &  n.s.                &  n.s.                &  n.s.                &  n.s.                &  n.s.                &  n.s.                &  n.s.                \\
HD\,158643     & $<$2 ($-$12)           & $<$1 ($-$12)           & $<$1 ($-$12)           & $<$3 ($-$12)           & $<$8 ($-$13)           & $<$2 ($-$11)           & $<$3 ($-$12)           \\
HD\,163296     &  n.s.                &  n.s.                &  n.s.                &  n.s.                &  n.s.                &  n.s.                & $<$9 ($-$12)           \\
HD\,169142     & 10$\pm$2 ($-$13)     & 442$\pm$9 ($-$14)    & 258$\pm$2 ($-$13)    & 9$\pm$3 ($-$13)      & 67$\pm$9 ($-$14)     & 729$\pm$2 ($-$13)    & 89$\pm$1 ($-$13)     \\
VV Ser       & $<$5 ($-$13)           & 25$\pm$4 ($-$14)     & 497$\pm$6 ($-$14)    & $<$1 ($-$13)           & $<$2 ($-$13)           & 153$\pm$3 ($-$13)    & 27$\pm$2 ($-$13)     \\
T CrA        &  n.s.                &  n.s.                &  n.s.                &  n.s.                &  n.s.                &  n.s.                & $<$9 ($-$13)           \\
HD\,179218     &  n.s.                &  n.s.                &  n.s.                &  n.s.                &  n.s.                &  n.s.                & 243$\pm$8 ($-$13)    \\
WW Vul       & $<$6 ($-$14)           & 12$\pm$1 ($-$14)     & 79$\pm$1 ($-$14)     & $<$7 ($-$14)           & $<$4 ($-$14)           & $<$6 ($-$13)           & $<$3 ($-$13)           \\
HD\,190073     & $<$2 ($-$13)           & $<$2 ($-$13)           & $<$2 ($-$13)           & $<$5 ($-$13)           & $<$2 ($-$13)           & $<$3 ($-$12)           & $<$2 ($-$12)           \\
LkHa 224     &  n.s.                &  n.s.                &  n.s.                &  n.s.                &  n.s.                &  n.s.                &  n.s.                \\
HD\,203024     & $<$2 ($-$13)           & $<$8 ($-$14)           & $<$1 ($-$13)           & $<$2 ($-$13)           & $<$1 ($-$13)           & $<$4 ($-$12)           & $<$1 ($-$12)           \\
\enddata
\tablecomments{The line fluxes and errors are in ergs
  cm$^{-2}$ s$^{-1}$, with $a\,(b)$ representing $a \times 10^{b}$. For
  non-detections, the 3-sigma upper limit is given. \mbox{n.s. = No}
  spectrum available at this wavelength.}
\end{deluxetable}

\begin{deluxetable}{lccccc}
\tabletypesize{\scriptsize}
\tablecaption{Line fluxes of the PAH features in the 10-14~$\mu$m
  range. \label{table_12_lineflux}}
\tablewidth{0pt}
\tablehead{
Target       &10.6&11.3&12.0&12.7&13.5
}
\startdata
HD\,31293      & $<$5 ($-$12)           & 11$\pm$3 ($-$12)     & 8$\pm$7 ($-$13)      & $<$2 ($-$12)           & 6$\pm$5 ($-$13)      \\
HD\,31648      & $<$2 ($-$12)           & $<$7 ($-$12)           & $<$2 ($-$12)           & $<$2 ($-$12)           & $<$1 ($-$12)           \\
HD\,34282      & 39$\pm$2 ($-$14)     & 554$\pm$4 ($-$14)    & 89$\pm$2 ($-$14)     & 138$\pm$2 ($-$14)    & 5$\pm$2 ($-$14)      \\
HD\,35187      & 11$\pm$8 ($-$14)     & 27$\pm$2 ($-$13)     & $<$3 ($-$13)           & 103$\pm$6 ($-$14)    & $<$2 ($-$13)           \\
HD\,35929      & $<$8 ($-$14)           & $<$1 ($-$13)           & $<$7 ($-$14)           & $<$6 ($-$14)           & $<$6 ($-$14)           \\
HD\,36112      & $<$7 ($-$13)           & $<$2 ($-$12)           & $<$5 ($-$13)           & $<$4 ($-$13)           & $<$3 ($-$13)           \\
HD\,244604     & $<$4 ($-$13)           & $<$7 ($-$13)           & $<$4 ($-$13)           & $<$3 ($-$13)           & $<$3 ($-$13)           \\
HD\,36917      & 27$\pm$3 ($-$14)     & 534$\pm$6 ($-$14)    & 59$\pm$3 ($-$14)     & 88$\pm$3 ($-$14)     & 20$\pm$3 ($-$14)     \\
HD\,37258      & $<$2 ($-$13)           & $<$9 ($-$13)           & $<$2 ($-$13)           & $<$2 ($-$13)           & $<$1 ($-$13)           \\
BF Ori       & $<$2 ($-$13)           & $<$9 ($-$13)           & $<$2 ($-$13)           & $<$2 ($-$13)           & $<$1 ($-$13)           \\
HD\,37357      & $<$3 ($-$13)           & $<$2 ($-$12)           & $<$2 ($-$13)           & $<$2 ($-$13)           & $<$2 ($-$13)           \\
HD\,37411      & $<$4 ($-$14)           & 65$\pm$7 ($-$14)     & $<$5 ($-$14)           & 22$\pm$2 ($-$14)     & $<$5 ($-$14)           \\
RR Tau       & 81$\pm$3 ($-$14)     & 658$\pm$7 ($-$14)    & 41$\pm$2 ($-$14)     & 175$\pm$2 ($-$14)    & 7$\pm$2 ($-$14)      \\
HD\,37806      & $<$1 ($-$12)           & $<$9 ($-$12)           & $<$1 ($-$12)           & $<$1 ($-$12)           & $<$9 ($-$13)           \\
HD\,38120      & $<$6 ($-$13)           & $<$1 ($-$12)           & $<$5 ($-$13)           & $<$6 ($-$13)           & $<$4 ($-$13)           \\
HD\,250550     & $<$1 ($-$13)           & $<$9 ($-$13)           & $<$1 ($-$13)           & $<$3 ($-$13)           & $<$1 ($-$13)           \\
HD\,259431     & 34$\pm$1 ($-$13)     & 308$\pm$3 ($-$13)    & 240$\pm$8 ($-$14)    & 73$\pm$1 ($-$13)     & 44$\pm$7 ($-$14)     \\
V590 Mon     & 51$\pm$5 ($-$14)     & 29$\pm$1 ($-$13)     & $<$1 ($-$13)           & 83$\pm$4 ($-$14)     & $<$1 ($-$13)           \\
HD\,50138      & $<$3 ($-$12)           & $<$3 ($-$11)           & $<$4 ($-$12)           & $<$1 ($-$11)           & $<$5 ($-$12)           \\
NX Pup       & $<$6 ($-$13)           & $<$5 ($-$13)           & $<$2 ($-$13)           & $<$3 ($-$13)           & $<$2 ($-$13)           \\
HD\,58647      & $<$2 ($-$13)           & $<$3 ($-$13)           & $<$2 ($-$13)           & $<$1 ($-$13)           & $<$1 ($-$13)           \\
HD\,72106S     & $<$3 ($-$13)           & 44$\pm$2 ($-$13)     & 43$\pm$9 ($-$14)     & 120$\pm$8 ($-$14)    & $<$2 ($-$13)           \\
HD\,85567      & $<$3 ($-$13)           & 39$\pm$2 ($-$13)     & 5$\pm$1 ($-$13)      & 74$\pm$9 ($-$14)     & $<$2 ($-$13)           \\
HD\,95881      & 14$\pm$1 ($-$13)     & 90$\pm$2 ($-$13)     & 8$\pm$1 ($-$13)      & 25$\pm$1 ($-$13)     & $<$3 ($-$13)           \\
HD\,97048      & 45$\pm$2 ($-$13)     & 633$\pm$4 ($-$13)    & 57$\pm$1 ($-$13)     & 1547$\pm$9 ($-$14)   & 18$\pm$1 ($-$13)     \\
HD\,98922      & 59$\pm$5 ($-$13)     & 18$\pm$1 ($-$12)     & 23$\pm$5 ($-$13)     & 75$\pm$8 ($-$13)     & $<$2 ($-$12)           \\
HD\,100453     & $<$7 ($-$13)           & 95$\pm$4 ($-$13)     & $<$7 ($-$13)           & 14$\pm$2 ($-$13)     & $<$5 ($-$13)           \\
HD\,100546     & $<$5 ($-$12)           & $<$6 ($-$11)           & $<$5 ($-$12)           & $<$1 ($-$11)           & $<$7 ($-$12)           \\
HD\,101412     & $<$2 ($-$13)           & 151$\pm$7 ($-$14)    & $<$1 ($-$13)           & 38$\pm$6 ($-$14)     & $<$1 ($-$13)           \\
HD\,104237     & $<$3 ($-$12)           & $<$9 ($-$12)           & $<$2 ($-$12)           & $<$2 ($-$12)           & $<$2 ($-$12)           \\
SS73 44      & $<$9 ($-$14)           & $<$2 ($-$13)           & $<$8 ($-$14)           & $<$9 ($-$14)           & $<$9 ($-$14)           \\
HD\,135344B    & $<$4 ($-$14)           & 260$\pm$3 ($-$14)    & 48$\pm$1 ($-$14)     & 60$\pm$1 ($-$14)     & 10$\pm$2 ($-$14)     \\
HD\,139614     & $<$3 ($-$13)           & $<$5 ($-$13)           & $<$3 ($-$13)           & $<$1 ($-$12)           & $<$2 ($-$13)           \\
HD\,141569     & 12$\pm$2 ($-$14)     & 83$\pm$3 ($-$14)     & 11$\pm$2 ($-$14)     & 29$\pm$1 ($-$14)     & 4$\pm$1 ($-$14)      \\
HD\,142666     & 2$\pm$1 ($-$13)      & 42$\pm$2 ($-$13)     & 82$\pm$9 ($-$14)     & 11$\pm$1 ($-$13)     & $<$2 ($-$13)           \\
HD\,142527     & $<$2 ($-$12)           & 161$\pm$8 ($-$13)    & $<$1 ($-$12)           & 25$\pm$4 ($-$13)     & $<$1 ($-$12)           \\
HD\,144432     & $<$1 ($-$12)           & $<$7 ($-$12)           & $<$9 ($-$13)           & $<$8 ($-$13)           & $<$7 ($-$13)           \\
HD\,144668     & $<$7 ($-$13)           & $<$9 ($-$12)           & $<$9 ($-$13)           & $<$1 ($-$12)           & $<$1 ($-$12)           \\
HD\,145263     & $<$7 ($-$14)           & $<$6 ($-$13)           & $<$6 ($-$14)           & $<$5 ($-$14)           & $<$5 ($-$14)           \\
Wray 15-1484 & 33$\pm$4 ($-$13)     & 212$\pm$6 ($-$13)    & 25$\pm$3 ($-$13)     & 52$\pm$3 ($-$13)     & 6$\pm$2 ($-$13)      \\
HD\,150193     & $<$2 ($-$12)           & $<$1 ($-$11)           & $<$2 ($-$12)           & $<$3 ($-$12)           & $<$3 ($-$12)           \\
HD\,152404     & $<$3 ($-$13)           & $<$2 ($-$12)           & $<$2 ($-$13)           & $<$1 ($-$13)           & $<$1 ($-$13)           \\
KK Oph       & $<$5 ($-$13)           & $<$9 ($-$13)           & $<$4 ($-$13)           & $<$4 ($-$13)           & $<$3 ($-$13)           \\
HD\,158643     & $<$1 ($-$12)           & $<$2 ($-$12)           & $<$1 ($-$12)           & $<$1 ($-$12)           & $<$8 ($-$13)           \\
HD\,163296     & $<$2 ($-$12)           & $<$2 ($-$11)           & $<$2 ($-$12)           & $<$2 ($-$12)           & $<$2 ($-$12)           \\
HD\,169142     & 88$\pm$7 ($-$14)     & 208$\pm$1 ($-$13)    & 206$\pm$6 ($-$14)    & 365$\pm$6 ($-$14)    & 9$\pm$5 ($-$14)      \\
VV Ser       & $<$3 ($-$13)           & 17$\pm$2 ($-$13)     & $<$3 ($-$13)           & $<$2 ($-$13)           & $<$2 ($-$13)           \\
T CrA        & $<$4 ($-$13)           & $<$1 ($-$12)           & $<$5 ($-$13)           & $<$9 ($-$13)           & $<$7 ($-$13)           \\
HD\,179218     & $<$4 ($-$12)           & $<$3 ($-$11)           & $<$1 ($-$12)           & $<$7 ($-$12)           & $<$4 ($-$12)           \\
WW Vul       & $<$2 ($-$13)           & $<$9 ($-$13)           & $<$1 ($-$13)           & $<$1 ($-$13)           & $<$9 ($-$14)           \\
HD\,190073     & $<$9 ($-$13)           & $<$2 ($-$12)           & $<$8 ($-$13)           & $<$7 ($-$13)           & $<$6 ($-$13)           \\
LkHa 224     & $<$6 ($-$13)           & $<$7 ($-$13)           & $<$3 ($-$13)           & $<$5 ($-$13)           & $<$3 ($-$13)           \\
HD\,203024     & $<$5 ($-$13)           & $<$9 ($-$13)           & $<$4 ($-$13)           & $<$2 ($-$12)           & $<$3 ($-$13)           \\
\enddata
\tablecomments{The line fluxes and errors are in ergs
  cm$^{-2}$ s$^{-1}$, with $a\,(b)$ representing $a \times
  10^{b}$. For 
  non-detections, the 3-sigma upper limit is given.}
\end{deluxetable}

\begin{deluxetable}{lcc}
\tabletypesize{\scriptsize}
\tablecaption{PAH-to-stellar and aliphatic-to-PAH
  luminosity ratios \label{table_LpahLstar}}
\tablewidth{0pt}
\tablehead{
Target       & L$_\mathrm{PAH}$/L$_\star$ &
 L$_\mathrm{ali}$/L$_\mathrm{PAH}$
}
\startdata
HD\,31293           &  $44  \pm 2$ ($-$4)    &   $13 \pm  2$ ($-$3)  \\
HD\,31648           &  $25  \pm 5$ ($-$4)    &   $28 \pm  7$ ($-$3)  \\
HD\,34282           &  $1554\pm 5$ ($-$5)    &   $7  \pm  1$ ($-$3)  \\
HD\,35187           &  $1228\pm 9$ ($-$6)    &   $15 \pm  1$ ($-$3)  \\
HD\,36112           &  $13  \pm 2$ ($-$4)    &   $28 \pm  7$ ($-$3)  \\
HD\,244604          &  $13  \pm 3$ ($-$4)    &   $<$2 ($-$3)         \\
HD\,36917           &  $602 \pm 5$ ($-$6)    &   $5  \pm  2$ ($-$3)  \\
BF Ori            &  $171 \pm 7$ ($-$6)    &   $<$1 ($-$2)         \\
HD\,37357           &  $9   \pm 1$ ($-$4)    &   $10 \pm  5$ ($-$3)  \\
HD\,37411           &  $34  \pm 1$ ($-$4)    &   $<$1 ($-$2)         \\
RR Tau            &  $2097\pm 8$ ($-$5)    &   $<$2 ($-$2)         \\
HD\,37806           &  $19  \pm 4$ ($-$4)    &   $21 \pm  5$ ($-$3)  \\
HD\,38120           &  $13  \pm 3$ ($-$4)    &   $3  \pm  1$ ($-$2)  \\
HD\,250550          &  $13  \pm 1$ ($-$4)    &   $34 \pm  9$ ($-$3)  \\
V590 Mon          &  $328 \pm 2$ ($-$4)    &   $<$1 ($-$1)         \\
HD\,58647           &  $83  \pm 2$ ($-$6)    &   $<$1 ($-$1)         \\
HD\,72106S          &  $248 \pm 3$ ($-$5)    &   $5  \pm  2$ ($-$3)  \\
HD\,85567           &  $88  \pm 1$ ($-$5)    &   $11 \pm  2$ ($-$3)  \\
HD\,95881           &  $552 \pm 4$ ($-$5)    &   $6  \pm  2$ ($-$3)  \\
HD\,97048           &  $901 \pm 3$ ($-$5)    &   $6  \pm  2$ ($-$3)  \\
HD\,100453          &  $47  \pm 1$ ($-$4)    &   $8  \pm  1$ ($-$3)  \\
HD\,101412          &  $190 \pm 7$ ($-$5)    &   $11 \pm  5$ ($-$3)  \\
HD\,135344B         &  $180 \pm 2$ ($-$5)    &   $32 \pm  8$ ($-$3)  \\
HD\,139614          &  $254 \pm 4$ ($-$5)    &   $23 \pm  6$ ($-$3)  \\
HD\,141569          &  $353 \pm 2$ ($-$6)    &   $<$1 ($-$1)         \\
HD\,142666          &  $194 \pm 2$ ($-$5)    &   $28 \pm  2$ ($-$3)  \\
HD\,142527          &  $295 \pm 5$ ($-$5)    &   $36 \pm  2$ ($-$3)  \\
HD\,144432          &  $29  \pm 1$ ($-$5)    &   $21 \pm  4$ ($-$2)  \\
HD\,152404          &  $106 \pm 5$ ($-$6)    &   $<$3 ($-$2)         \\
HD\,169142          &  $605 \pm 2$ ($-$5)    &   $11 \pm  3$ ($-$3)  \\
VV Ser            &  $411 \pm 6$ ($-$5)    &   $<$1 ($-$1)         \\
WW Vul            &  $347 \pm 7$ ($-$6)    &   $<$3 ($-$2)         \\
\enddata
\tablecomments{Only targets with a full 5--37-\mic IRS spectrum and
  with detected PAH emission are included. $a\,(b)$ represents $a
  \times 10^{b}$.}
\end{deluxetable}

\begin{deluxetable}{lllccccccc}
\tablecaption{Correlations \label{table_corr}}
\tablewidth{0pt}
\tablehead{
\multicolumn{2}{c}{Parameters} & & p-value
}
\startdata
Cen 7.8\,\mic  & \Teff         & \down & $4 \times 10^{-3}$  \\
PP 7.8\,\mic   & \Teff         & \down & $2 \times 10^{-3}$  \\
FWHM 7.8\,\mic & \Teff         & \down & $6 \times 10^{-3}$  \\
Cen 6.2\,\mic  & \Teff         & \down & $4 \times 10^{-3}$  \\
PP 6.2\,\mic   & \Teff         & \down & $5 \times 10^{-4}$  \\
LF 8.6/6.2     & \Teff         & \up   & $2 \times 10^{-3}$  \\
LF 8.6/7.8     & \Teff         & \up   & $5 \times 10^{-3}$  \\
\LaliLpah      & \Teff         & \down & $3 \times 10^{-2}$  \\
LF 8.6/ali     & \Teff         & \up   & $3 \times 10^{-3}$  \\
FWHM 7.8\,\mic & Cen 7.8\,\mic & \up   & $1 \times 10^{-3}$  \\
LF 8.6/6.2     & Cen 7.8\,\mic & \down & $1 \times 10^{-3}$  \\
LF 8.6/7.8     & Cen 7.8\,\mic & \down & $4 \times 10^{-4}$  \\
\LaliLpah      & Cen 7.8\,\mic & \up   & $4 \times 10^{-3}$  \\
LF 8.6/ali     & Cen 7.8\,\mic & \down & $3 \times 10^{-3}$  \\
LF 8.6/6.2     & \LaliLpah     & \down & $1 \times 10^{-2}$  \\
LF 8.6/7.8     & \LaliLpah     & \down & $4 \times 10^{-3}$  \\
\enddata
\tablecomments{A summary of the correlations indicating the
dependence of the hydrocarbon chemistry upon the stellar radiation
field. The arrows indicate a positive or negative correlation. The
p-value is the probability of a non-correlation. Cen: centroid
position; PP: peak position; LF: line flux ratio; FWHM: full width at
half maximum; ali: the aliphatic 6.8 and 7.2-\mic CH features.}
\end{deluxetable}

\end{document}